\documentclass[aps,prd,amsmath,floats,floatfix,twocolumn,nofootinbib,superscriptaddress,showpacs]{revtex4-1}
\usepackage{graphicx}
\usepackage{multirow}
\usepackage{color}
\usepackage{latexsym}
\usepackage{amsfonts}
\usepackage{amsmath}
\usepackage[colorlinks]{hyperref}
\hypersetup{
  colorlinks=true,        % false: boxed links; true: colored links
  linkcolor=blue,         % color of internal links
  citecolor=cyan,         % color of links to bibliography
}
\graphicspath{{figures/}} % Location of the graphics files

\newcommand{\be}{\begin{eqnarray}}
\newcommand{\ee}{\end{eqnarray}}

\begin{document}

	\title{Comparing spin supplementary conditions for particle motion around traversable wormholes}

\author{Carlos~A.~Benavides-Gallego}
	\email{cabenavidesg@sjtu.edu.cn}
    \affiliation{School of Aeronautics and Astronautics, Shanghai Jiao Tong University, Shanghai 200240, PRC.}
	\affiliation{School of Physics and Astronomy, Shanghai Jiao Tong University, 800 Dongchuan Road, Minhang, Shanghai 200240, PRC.}
    \affiliation{Shanghai Frontiers Science Center of Gravitational Wave Detection, 800 Dongchuan Road, Minhang, Shanghai 200240, People's Republic of China}

\author{Jose Miguel Ladino}
    \email[]{jmladinom@unal.edu.co}% Your name
    \affiliation{Universidad Nacional de Colombia, Sede Bogotá. Facultad de Ciencias. Observatorio Astronómico Nacional. Ciudad Universitaria. Bogotá, 111321, Colombia.}

\author{Eduard Larra\~{n}aga}
    \email[]{ealarranga@unal.edu.co}% Your name
    \affiliation{ Universidad Nacional de Colombia, Sede Bogotá. Facultad de Ciencias. Observatorio Astronómico Nacional. Ciudad Universitaria. Bogotá, 111321, Colombia. }

	\date{\today}

	\begin{abstract}
        The Mathisson-Papapetrou-Dixon (MPD) equations describe the motion of spinning test particles in the pole-dipole approximation. It is well-known that these equations, which couple the Riemann curvature tensor with the antisymmetric spin tensor $S^{\alpha\beta}$, together with the normalization condition for the four-velocity, is a system of eleven equations relating fourteen unknowns. To ``close'' the system, it is necessary to introduce a constraint of the form $V_\mu S^{\mu \nu} = 0$, usually known as the spin supplementary condition (SSC), where $V_\mu$ is a future-oriented reference vector satisfying the normalization condition $V_\alpha V^\alpha = -1$. There are several SSCs in the literature. In particular, the Tulzcyjew-Dixon, Mathisson-Pirani, and  Ohashi-Kyrian-Semerák are the most used by the community. From the physical point of view, choosing a different SSC (a different reference vector $V^\mu$) is equivalent to fixing the centroid of the test particle. In this manuscript, we compare different SSCs for spinning test particles moving around a Morris-Thorne traversable wormhole. To do so, we first obtain the orbital frequency and expand it up to third-order in the particle's spin; as expected, the zero-order coincides with the Keplerian frequency, the same in all SSCs; nevertheless, we found that differences appear in the second order of the expansion, similar to the Schwarzschild and Kerr black holes. We also compare the behavior of the innermost stable circular orbit (ISCO). Since each SSC is associated with a different centroid of the test particle, we analyze (separately) the radial and spin corrections for each SSC. We found that the radial corrections improve the convergence, especially between Tulzcyjew-Dixon and Mathisson-Pirani SSCs. In the case of Ohashi-Kyrian-Semerák, we found that the spin corrections remove the divergence for the ISCO and extend its existence for higher values of the particle's spin   
	\end{abstract}
	
	\maketitle
	
%%%%%%%%%%%%%%%%%%%%%%%%%%%%%%%%%%%%%%%%%%%%%%%%%%%%%%%%%%%%%%%%%%%%%%%%%%%%%%%%%%%% Section I %%%%%%%%%%%%%%%%%%%%%%%%%%%%
%%%%%%%%%%%%%%%%%%%%%%%%%%%%%%%%%%%%%%%%%%%%%%%%%%%%%%%%%%%%%%
    \section{INTRODUCTION} \label{sec:INTRODUCTION}

    The dynamics of extended bodies is a crucial problem in any theory of gravity~\cite{Almonacid:2013xnu, Almonacid:2017gbz}. In the particular case of general relativity (GR), the problem has been investigated since Einstein's theory was published, and two approaches have been developed throughout history~\cite{MPD03}. In the first approach, based on Einstein's philosophy, bodies were considered as a set of elementary particles~\cite{Einstein:1938yz, Einstein:1940mt, Fock1964}. In the second approach, on the other hand, bodies were treated using multipole moments and assuming them small so that they do not affect the spacetime background. This approach corresponds to the work of Mathisson, Papapetrou, Tulczyjew, and Dixon~\cite{MPD03, MPD01, MPD01new, MPD02, Tulczyjew1959}.    

    In contrast to particles without internal structure, extended bodies do not follow a geodesic\footnote{ Initially, in the case of point particles, this was a postulate of the theory, but later, Einstein showed it was a consequence of his field equations~\cite{EinsteinI1927, EinsteinII1927}.}. Therefore, obtaining the equations of motion is more challenging mainly because GR describes physical bodies using a four-dimensional region of spacetime, which is not independent of the relationship between gravity and geometry. In this sense, it is troublesome to establish the laws of motion of the body since its trajectory (world-line) is affected by itself, the gravitational field, and the geometry, which depends on the momentum-energy distribution~\cite{Almonacid:2013xnu}.
    
    In particular, if one considers extended bodies as a set of elementary particles (Einstein's point of view), it is clear that this approach generates problems when facing the question of the motion of celestial bodies because it would be necessary treating them as an assembly of elementary particles. According to Dixon, this requires ``\textit{a general relativistic version of statistical mechanics, a formidable task}''~\cite{Dixon2015}. Therefore, from the astronomical point of view, a more suitable approach is to assume sufficiently small individual bodies so that one can treat them as particles~\cite{Robertson1937}. That, however, required answering two important questions~\cite{Dixon2015}. First, what point to choose to describe the position of the body? Second, how to describe its structure? Mathisson was the first to address these questions in 1937. To solve them, he started by selecting an arbitrary world-line within the body to characterize its motion and position and considering the energy-momentum tensor as an infinite set of multipole moments. In this way, the covariant conservation of the energy-momentum tensor becomes a set of equations representing the evolution of the multipole moments. All these considerations are contained in what Mathisson named ``\textit{the gravitational skeleton of a body}''~\cite{Dixon2015}. Hence, chronologically speaking, he was the first to introduce the concepts of multipole moments and multipole particles in GR~\cite{MPD01}.

    In 1951, Papapetrou also considered an extended test body as described by an energy-momentum tensor, $T^{\mu\nu}$, and he defined in a non-covariant way its multipole moments~\cite{MPD02}. Then, using the conservation equation $\nabla_\beta T^{\alpha\beta}=0$, he obtained the equations for a line inside the world-tube of the body under the assumption that all the moments higher than the dipole moment can be neglected. Furthermore, similarly to Mathisson, Papapetrou imposed the supplementary condition $V_\alpha S^{\alpha\beta}=0$ to make the equations fully determinate; however, in contrast to Mathisson, where $V_\alpha$ was the four-velocity, $u_\alpha$, in Papapetrou's development, $V_\alpha$ corresponds to a vector field determined by the metric and not related to the body under consideration. Back then, the advantage of considering such a vector lay in avoiding the nonphysical helical motions allowed by Mathisson's equations. Today, thanks to the work of Costa et al., we know that helical motions are perfectly valid and physically equivalent to the dynamics of a spinning body; the only difference is the choice of the representative point of the particle, a gauge choice~\cite{Costa:2011zn}.
    
    In 1959, Tulczyjew simplified Mathisson's theory by describing the particle not by a singularity in a linearized disturbance but by a singular energy-momentum tensor~\cite{Dixon2015}. He obtained the same equations as Mathisson without imposing the supplementary condition $u_\alpha S^{\beta\alpha}=0$, and showing that the momentum vector of the particle, $p^\alpha$, and its four-velocity, $u^\alpha$, are not parallel~\cite{Tulczyjew1959}. In a succeeding work~\cite{BTulczyjew1962}, Tulczyjew considered Papapetrou's approach in a more covariant form by introducing the world-line of the center of mass, related to the condition $p_\beta S^{\alpha\beta}=0$~\cite{Dixon2015, BTulczyjew1962}. Finally, in 1964, Dixon proposed a new treatment to the problem of extended bodies in GR, in which he included the effect of the electromagnetic field and a covariant definition of the center of mass. This allowed him to obtain the equations of motion in the pole-dipole approximation~\cite{Dixon:1970zza, Dixon:1970zz}. According to Dixon, these equations are applicable to macroscopic bodies such as a planet orbiting the Sun but not to bodies of atomic scales since GR breaks down when quantum phenomena become important.

    Nowadays, the motion of extended bodies in the pole-dipole approximation is described by the Mathisson-Papapetrou-Dixon (MPD) equations \cite{MPD01, MPD01new, MPD02, MPD03},
    \begin{align}
        \frac{D p^\mu}{d\lambda} = &-\frac{1}{2} R^\mu_{\nu \rho \sigma} u^\nu S^{\rho \sigma} \label{eq:MPD1}\\
        \frac{D S^{\mu \nu}}{d\lambda} =& p^\mu u^\nu - u^\mu p^\nu\label{eq:MPD2},
    \end{align}
    where the spinning test particle is characterized by a velocity vector, $u^\mu$, and momentum, $p^\mu$, in a curved spacetime background with the Riemann tensor defined as
    \begin{equation}
    \label{s1e3}
    R^\mu_{\nu \kappa \lambda}=\Gamma^\mu_{\kappa\alpha}\Gamma^\alpha_{\lambda\nu}-\Gamma^\mu_{\lambda\alpha}\Gamma^\alpha_{\kappa\nu}-\partial_\lambda\Gamma^\mu_{\kappa\nu}+\partial_\kappa\Gamma^\mu_{\lambda\nu}.
    \end{equation}
    In Eqs.~(\ref{eq:MPD1}) and (\ref{eq:MPD2}), $\frac{D}{d\lambda}\equiv u^\mu \nabla_\mu$ is the absolute derivative and $\lambda$ is an affine parameter. The antisymmetric tensor, $S^{\mu \nu} = - S^{\nu \mu}$, is related to the particle's spin. 
    
    As mentioned above, note that in contrast to non-spinning test particles, where the absolute derivative of the four-momentum $p^\mu$ vanishes (the geodesic equation), spinning test particles follow an equation of motion coupled to $S^{\mu \nu}$ and $R^\mu_{\nu \kappa \lambda}$. From the physical point of view, this means that spinning test particles do not follow a geodesic. As a consequence, $u^\mu$ and $p^\mu$ are not parallel, and one needs to introduce two concepts of mass, the dynamical and the kinematical rest masses, defined by 
    \begin{align}
    \mu^2 = & - p_\alpha p^\alpha,\\
    m = & - p_\alpha u^\alpha.
    \end{align}
    The non-parallel behavior between the  velocity and the momentum is obtained by contracting the second MPD Eq.~ (\ref{eq:MPD2}) with the velocity and taking into account the normalization condition $u_\alpha u^\alpha = -1$. Hence, one obtains the following relation
    \begin{equation}
    p^\alpha = m u^\alpha + p^\alpha_{\text{hidden}},
    \end{equation}
    where the term $p^\alpha_{\text{hidden}} = u_\beta \frac{D S^{\alpha \beta}}{d\lambda}$ is called the \textit{hidden momentum}, which depends on the behavior of the spin tensor along the trajectory.
    
    The MPD system of Eqs.~ (\ref{eq:MPD1}) and (\ref{eq:MPD2}), together with the normalization condition for the velocity, is a set of eleven equations relating fourteen unknown variables: $[p^\alpha, u^\alpha, S^{\alpha \beta}]$. To ``close'' this system, it is necessary to introduce a constraint equation, usually known as the spin supplementary condition (SSC); one  can find several SSCs in the literature represented by the general form 
    \begin{equation}
    V_\mu S^{\mu \nu} = 0, \label{eq:generalSSC}
    \end{equation}
    where $V^\mu$ is a future-oriented reference vector satisfying the normalization condition $V_\alpha V^\alpha = -1$. As mentioned above, while developing the spinning test particle dynamics, Mathisson, Papapetrou, Pirani, and Tulczyjew used different SSCs~\cite{MPD03, Tulczyjew1959,Pirani1956, Pirani2009}. Other examples are the the Ohashi-Kyrian-Semeràk (OKS)~\cite{OKS01, OKS02, OKS03}, Corinaldesi-Papapetrou~\cite{Corinaldesi:1951pb} and the Newton-Wigner~\cite{Newton:1949cq} SSCs. From the physical point of view, choosing a particular reference vector $V^\mu$ corresponds to fixing the centroid of the body.
    
    The MPD equations have been used widely in the literature, see \cite{Wald:1972sz, Barker:1979dy, Tod:1976ud, Carmeli:1976mq, Hojman:1976kn, Semerak:1999qc, Kyrian:2007zz, Shibata:1993uk, Mino:1995fm, Costa:2017kdr} and references therein. For example, The effects of the spin-curvature interaction were discussed by Wald (1972) and Barker $\&$ O'Connell (1979), Refs.~\cite{Wald:1972sz} and \cite{Barker:1979dy}, respectively. The motion of spinning test particles in the field of a black hole was investigated by K. P. Tod et al. in Ref.~\cite{Tod:1976ud}, while the dynamics in Vaidya's radiating metric and the Kerr-Newman spacetime were considered in Refs.~\cite{Carmeli:1976mq} and \cite{Hojman:1976kn}, respectively. Latter, Semeràk, and K. Kyrian and Semeràk numerically solved the MPD equations to investigate the trajectories of spinning particles in the Kerr black hole using the TD SSC~\cite{Semerak:1999qc, Kyrian:2007zz}. There, when the pole-dipole approximation is considered, the author found that no significant spin effects are expected if one considers astrophysical scenarios. Nevertheless, during the inspiral of a spinning particle onto a rotating compact body, important effects may occur that would modify the gravitational waves generated by the system. The gravitational wave generated by a spinning test particle falling or orbiting a black hole was investigated by Masaru Shibata and Yasushi Mino et al. in Refs.~\cite{Shibata:1993uk} and \cite{Mino:1995fm}, respectively. 

    Recently, the MPD equations have been used to investigate the motion of spinning test particles in different spacetimes~\cite{Plyatsko:2013xza, Hackmann:2014tga, Jefremov:2015gza, Harms:2016ctx, Zhang:2017nhl, Toshmatov:2019bda, Conde:2019juj, Larranaga:2020ycg, Timogiannis:2021ung, Benavides-Gallego:2021lqn, Abdulxamidov:2022ofi, Timogiannis:2022bks, Ladino:2022aja, Li:2022sjb}. The Kerr spacetime was considered in Refs.\cite{Plyatsko:2013xza, Hackmann:2014tga, Jefremov:2015gza, Harms:2016ctx}, in the last reference the authors also investigated the motion of spinning test particle in the Schwarzschild case. The properties of the innermost circular orbits (ISCO) for spinning test particles were investigated using the Kerr-Newman background in Ref.~\cite{Zhang:2017nhl}. The $\gamma$-metric, the Maxwell dilaton black hole, the charged Hayward black hole background, and the rotating black hole surrounded by the perfect fluid dark matter were studied in Refs.~\cite{Toshmatov:2019bda}, \cite{Conde:2019juj}, \cite{Larranaga:2020ycg} and \cite{Li:2022sjb}, respectively. The dynamics of spinning test particles in a quantum-improved rotating black Hole (RBH) spacetime was examined by M.~Ladino and E.~Larranaga in Ref.~\cite{Ladino:2022aja}. In the case of wormholes, the motion of spinning test particles was investigated in Refs.~\cite{Benavides-Gallego:2021lqn} and \cite{Abdulxamidov:2022ofi} using the Morris-Thorne non-rotating traversable wormhole~\cite{Morris1988} and the rotating traversable wormhole obtained by Teo~\cite{Teo:1998dp}, respectively. 
    
    Finally, in Refs.~\cite{Timogiannis:2021ung} and \cite{Timogiannis:2022bks}, the authors examine whether the equatorial circular orbits around a massive black hole are affected when the particle's centroid (associated with the SSC) changes to another centroid (different SSC). To do so, the authors established an analytical algorithm to obtain the orbital frequency of a spinning body moving around an arbitrary stationary, axisymmetric spacetime, focusing on three SCCs: the Tulzcyjew-Dixon (TD), the Mathisson-Pirani (MP), and the Ohashi-Kyrian-Semer\`ak SSCs (OKS). In the case of the Schwarzschild black hole, they investigated the discrepancies in the orbital frequency employing a power series expansion of the spin for each SSC, imposing corrections to improve the convergence between the SSC. They found that the shifting from one circular equatorial orbit to another, the coincidence between the SSCs only holds up to the third order in the orbital frequency. 

    In the Kerr spacetime, on the other hand, the authors considered the convergence of the orbital frequency for prograde and retrograde equatorial circular orbits. Following a similar approach used for the Schwarzschild case, they expanded the orbital frequencies in powers of the particle's spin for the same SSCs, i.e., the TD, MP, and OKS SSCs. The authors also introduced a novel method to compute the ISCO radius for any SCC. Similar to the Schwarszchild case, there is a convergence in the power series of the frequencies of the SSCs. However, there is a limit to this convergence because, in the spinning body approximation, one only considers the first two multipoles (pole-dipole) of the body and ignores the higher ones.   
    
    In this work, we follow Refs.~\cite{Timogiannis:2021ung, Timogiannis:2022bks} to compare the orbital frequency of equatorial circular orbits in the background of a traversable non-rotating wormhole. We consider three of the most well-known SSCs: the TD~\cite{MPD03, Tulczyjew1959}, MP~\cite{Pirani1956, Pirani2009}, and the OKS~\cite{OKS01, OKS02, OKS03} supplementary conditions. This work is organized as follows. In Sec.~\ref{sec:2}, we roughly describe the Morris-Thorne traversable wormhole. In Sec.~\ref{sec:3}, we obtain the orbital frequencies for equatorial circular orbits using the analytical algorithm proposed in Ref.~\cite{Timogiannis:2021ung} and obtain the analytical expressions for the Morris-Thorne wormhole using different centroids. Then, in Sec.~\ref{sec:4}, we discuss the orbital parameters; i.e., the orbital frequency and the ISCO radius, using different SSCs. In Sec.~\ref{sec:5}, we use the centroid corrections to explain the differences between the SSCs. Here we consider the corrections to the position of the centroid and corrections to the spin. Finally, in Sec.~\ref{sec:6}, we review and conclude our work.
    
    In the manuscript, we use dimensionless units, the Riemann curvature tensor is defined as Eq.~(\ref{s1e3}), and the metric has the signature $(-,+,+,+)$.
	
%%%%%%%%%%%%%%%%%%%%%%%%%%%%%%%%%%%%%%%%%%%%%%%%%%%%%%%%%%%%%%%%%%%%%%%%%%%%%%%%%%%%% Section II %%%%%%%%%%%%%%%%%%%%%%%%%%%%
%%%%%%%%%%%%%%%%%%%%%%%%%%%%%%%%%%%%%%%%%%%%%%%%%%%%%%%%%%%%%%
\section{MORRIS-THORNE WORMHOLES} \label{sec:2}
    
\begin{figure}[!htb]
    \centering
    \includegraphics[width=1\linewidth]{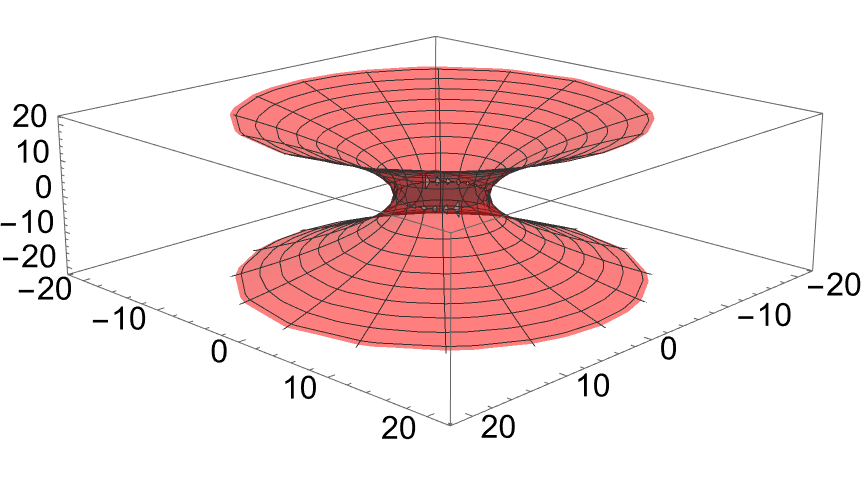}
    \caption{Wormhole spacetime embedded in a three-dimensional space. We consider $b_0=4$. \label{embedding}}
    \end{figure}

    The Morris-Thorne traversable wormhole is a spherically symmetric spacetime given by the line element~\cite{Morris1988, Visser2002} \begin{equation}
    d s^2=-e^{2 \Phi(r)} d t^2+\frac{d r^2}{1-b(r)}+r^2\left(d \theta^2+\sin ^2 \theta d \varphi^2\right),\label{eq:WHmetric}
    \end{equation}
    where $\Phi(r)$ and $b(r)$ are arbitrary functions of the radial coordinate $r$ known as the ``\textit{redshift function}'' and the ``\textit{shape function}'', respectively. The fact that Eq.~(\ref{eq:WHmetric}) represents a wormhole with a throat connecting two different regions of the spacetime can be easily depicted by embedding the line element in a three-dimensional space at a fixed time slice $t$, see Fig.~\ref{embedding}.

    %%%%%%%%%%%%% Figure 1 %%%%%%%%%%%%%%%%
    %%%%%%%%%%%%%%%%%%%%%%%%%%%%%%%%%%%%%%%
    
    The metric of Eq.~ (\ref{eq:WHmetric}) was obtained assuming first the wormhole's spacetime as spherically symmetric, then, via the field equations, computing the corresponding energy-momentum tensor. In contrast to black holes, wormholes do not have an event horizon or singularities; this means the redshift function $\Phi(r)$ is everywhere finite, and the spacetime has a throat connecting two asymptotically flat regions of the same universe.

    On the other hand, a wormhole is traversable if the tidal gravitational forces experienced by any traveler are bearable small~\cite{Morris1988}. Moreover, the time needed to cross the wormhole must be finite and reasonably small. From the physical point of view, this means that the proper time measured by a traveler and the observers outside the wormhole must be finite and small. Hence, in the case of a zero-tidal-force solution, the redshift and the shape functions have the form~\cite{Bambi:2013jda}
    \begin{equation}
    \label{S2e2}
    \Phi(r)=-\frac{b_0}{r} \quad \text { and } \quad b(r)=\left( \frac{b_0}{r} \right)^\gamma,
    \end{equation}
    where $b_0$ is the throat of the wormhole, usually associated with the wormhole's mass. We use these functions to compare the different SSCs, focusing on the values $\gamma=1$ and $\gamma=2$.
    
    %%%%%%%%%%%%%%%%%%%%% Figure 2%%%%%%%%%
    \begin{figure}[!htb]
    \centering
    \includegraphics[width=1\linewidth]{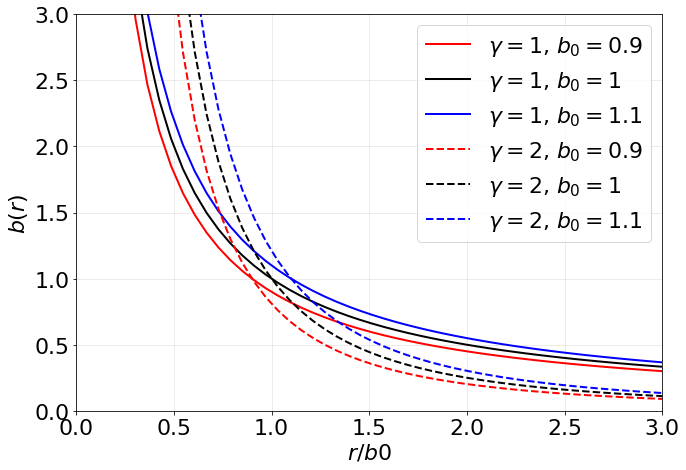}
    \caption{Plot of b(r), the ``\textit{shape function}'', as function of r, for different values of $b_0$ and for both cases of wormhole  solution $\gamma=1,2$. \label{shape}}
    \end{figure}
    %%%%%%%%%%%%%%%%%%%%%%%%%%%%%%%%%%%%%%%
    
    Fig.~\ref{shape} illustrates the behavior of the shape function $b(r)$ as a function of $r$ for various values of the wormhole throat $b_0$, and both $\gamma=1$ and $\gamma=2$ cases of the solution. Here we provide a preliminary comparison of the properties of the two wormhole solutions. The figure shows that as the wormhole throat $b_0$ increases, the shape function, $b(r)$, also increases in both solutions. Furthermore, the value of $b(r)$ for the $\gamma=2$ solution is asymptotically smaller and closer to $r=0$ than that of the $\gamma=1$ solution. Note that the shape function $b(r)$ has the same value for all cases at the wormhole throat point $b_0$.

    The properties mentioned above are essential for traversable wormholes, what Morris and Thorne refer to as \textit{basic wormhole criteria}. These criteria are deeply related to the form of the energy-momentum tensor, which depends on the matter and fields that generate wormholes. Even though the energy-momentum tensor is not physically reasonable since exotic mater (negative energy density) is required to create the wormhole's spacetime curvature at its throat, it is possible to tune the wormhole's parameters to make its constitution material compatible with the form of matter allowed by the laws of physics~\cite{Morris1988}.

%%%%%%%%%%%%%%%%%%%%%%%%%%%%%%%%%%%%%%%%%%%%%%%%%%%%%%%%%%%%%%%%%%%%%%%%%%%%%%%%%%%% Section III %%%%%%%%%%%%%%%%%%%%%%%%%%%%
%%%%%%%%%%%%%%%%%%%%%%%%%%%%%%%%%%%%%%%%%%%%%%%%%%%%%%%%%%%%%%
    \section{CIRCULAR EQUATORIAL ORBITS} \label{sec:3}
    Without loss of generality,  we can consider a spherically symmetric and static spacetime, described by the following line element:
    \begin{equation}
    ds^2 = g_{tt} dt^2 + g_{rr} dr^2 + g_{\theta \theta}d\theta^2 + g_{\varphi \varphi}d\varphi^2.
    \label{eq:SSSmetric}
    \end{equation}

    For circular equatorial orbits, we fix the spatial coordinates as $r=\text{costant}$, $\theta=\frac{\pi}{2}$, and $\varphi=\Omega t$, where $\Omega=u^\varphi / u^t$ is the orbital frequency of the spinning test body. To maintain circularity in all the SSCs, the radial and polar four-velocity and four-momentum components must vanish; i.e., $u^r=u^\theta=0$ and $p^r=p^\theta=0$, respectively. Moreover, the normalization condition of the four-velocity implies
    \begin{equation}
    u^t=\frac{1}{\sqrt{-g_{tt}-g_{\varphi \varphi} \Omega^2}}. \label{eq:utorbitalfreq}
    \end{equation}
    Considering a spinning test body moving in the equatorial plane, with a spin $S$ aligned (or anti-aligned) with the total angular momentum $J_z$ (perpendicular to the equatorial plane), it follows that the spin four-vector $S_\mu$ takes the form:
    \begin{equation}
    S_\mu:=-\frac{1}{2} \epsilon_{\mu \nu \rho \sigma} V^\nu S^{\rho \sigma},
    \end{equation}
    and
    \begin{equation}
    S^{\rho \sigma}=-\epsilon^{\rho \sigma \nu \kappa} S_\nu V_\kappa.
    \end{equation}
    Therefore, we can write $S^\mu \equiv S^\theta \delta_\theta^\mu$, and then the magnitude of the spin would be $S= \pm \sqrt{g_{\theta \theta}} S^\theta$. Hence, introducing the SSC in the general form of Eq.~ (\ref{eq:generalSSC})  and considering the definition of the spin angular momentum as
    \begin{equation}
    S^2=\frac{1}{2} S^{\mu v} S_{\mu \nu},
    \label{eq14}
    \end{equation}
    the only non-vanishing components of the spin tensor are
    \begin{align}
    \begin{cases}
    & S^{t r}=-S^{r t}=-S \sqrt{-\frac{g_{\theta \theta}}{g}} V_\varphi, \\
    & S^{r \varphi}=-S^{\varphi r}=-S \sqrt{-\frac{g_{\theta \theta}}{g}} V_t,
    \end{cases} \label{eq:SpinTensorComponents}
    \end{align}
    with $g=g_{tt} g_{rr} g_{\theta \theta}g_{\varphi \varphi}$ the determinant of the metric. To relate the non-vanishing components of the spin tensor with the conserved quantities, we use the Killing equation
    \begin{equation}
    C_\xi=\xi^\mu P_\mu-\frac{1}{2} S^{\mu \nu} \nabla_\nu \xi_\mu;
    \end{equation}
    where $\xi^\mu$ is a Killing vector field associated with the conserved quantity. 

    In the case of a static and spherically symmetric spacetime, represented by the line element of Eq.~\eqref{eq:SSSmetric}, two Killing vectors exist. These killing vectors, given by $\xi_{(t)}^\mu=\delta_t^\mu $ and  $ \xi_{(\varphi)}^\mu=\delta_\varphi^\mu$, are related to the conservation of both the energy $C_{(t)}=-E$ and the $z$ component of the total angular momentum of the spinning test particle $C_{(\varphi)}=J_z$, respectively. Using the Eqs.~\eqref{eq:SpinTensorComponents}, these conserved quantities can be expressed as~\cite{Timogiannis:2021ung} 
    \begin{align}
    \begin{cases}
    & E = -p_t- \partial_r g_{tt}  \frac{S}{2} \sqrt{-\frac{g_{\theta \theta}}{g}} V_\varphi, \\
    & J_z = p_\varphi- \partial_r g_{\varphi \varphi} \frac{S}{2} \sqrt{-\frac{g_{\theta \theta}}{g}} V_t.
    \end{cases} \label{eq:ConservedQuantities}
    \end{align}
    Hence, with the help of the above considerations and results, the MPD equations for equatorial circular orbits reduce to~\cite{Timogiannis:2021ung}
    \begin{align}
    \Gamma^r_{\rho \sigma} u^\rho p^\sigma= &- \frac{1}{2} R^r_{\nu \rho \sigma} u^\nu S^{\rho \sigma} \label{eq:circularMPDequation1} \\
    \Gamma^t_{\rho r} u^\rho S^{r \varphi} + \Gamma^\varphi_{\rho r} u^\rho S^{tr} =& p^t u^\varphi - p^\varphi u^t. \label{eq:circularMPDequation2}
    \end{align}
    In the following subsections, we apply the MPD Eqs.~\eqref{eq:circularMPDequation1} and \eqref{eq:circularMPDequation2} to different SSCs.

%%%%%%%%%%%%%%%%%%%%%%%%%%%%%%% TD SSC %%%%%%%%%%%%%%%%%%%%%%%%%%%%%%%%%%
    \subsection{Tulzcyjew-Dixon SSC} \label{sec:3-1}

    In the case of the Tulzcyjew-Dixon SSC (TD-SSC), the reference four-vector, $V^\mu$, is substituted by $p^\mu/\mu$. Therefore, after replacing the non-zero components of the spin tensor $S^{\mu\nu}$ shown in Eq.~(\ref{eq:SpinTensorComponents}) into Eqs.~(\ref{eq:circularMPDequation1}) and (\ref{eq:circularMPDequation2}), one obtains~\cite{Timogiannis:2021ung}   
    \begin{equation}
    \label{S3Ae1}
    \begin{aligned}
    \frac{p_t}{p_\varphi}&=f_1(r,\Omega;S),\\
    \frac{p_t}{p_\varphi}&=f_2(r,\Omega;S).
    \end{aligned}
    \end{equation}
    After equating $f_1$ and $f_2$, it is possible to obtain a second-order polynomial expression $\Omega$~\cite{Timogiannis:2021ung}
    \begin{equation}
    \rho_2\Omega^2+\rho_1\Omega+\rho_0=0, \label{eq:TDOmegapoly}
    \end{equation}
    from which~\cite{Timogiannis:2021ung}
    \begin{equation}
    \label{S3Ae3}
    \Omega_\pm=\frac{-\rho_1\pm\sqrt{\rho^2_1-4\rho_2\rho
    _0}}{2\rho_2},
    \end{equation}
    where $\Omega_+$ corresponds to the circular orbits in which $E>0$ and $J_z>0$, while $\Omega_-$ to $E>0$ and $J_z<0$. This is the usual convention valid for all the SSC considered in this paper. In the case of the Morris-Thorne wormhole~(\ref{eq:WHmetric}) with $\gamma=1$, $\rho_2$, $\rho_1$ and $\rho_0$ take the form
    \begin{equation}
    \label{S3Ae4}
    \begin{aligned}
    \rho_2&=\frac{1}{2}r e^{-\frac{2 b_0}{r}} \left(b_0 S^2 r- 2\mu ^2 r^4\right),\\
    %%%%%%%%
    \rho_1&=\frac{1}{2} b_0 \mu S e^{-\frac{3 b_0}{r}} \sqrt{r\left(r-b_0\right)} \left(2 b_0-5 r\right),\\
    %%%%%%%%
    \rho_0&=\frac{b_0 e^{-\frac{4 b_0}{r}} \left(b_0 S^2 \left(-7 b_0 r+2 b_0^2+4 r^2\right)+2 \mu ^2 r^5\right)}{2 r^3},\\
    \end{aligned}
    \end{equation}
    where we used Eq.~(1) in the appendix A of Ref.~\cite{Timogiannis:2021ung}. The corresponding expressions for $\gamma=2$ are given in Appendix \ref{sec:appendixTD}. 

    \begin{widetext}
    On the other hand, from the definition of the dynamical mass ($\mu:=\sqrt{-p_\nu p^\nu}$), the expressions, for $p^t$ and $p^\varphi$ for the Morris-Thorne wormhole are given by the
    \begin{equation}
        \label{S3Ae5}
        \begin{aligned}
        p^\varphi&=\frac{\mu}{r\sqrt{r^4 e^{\frac{2 b_0}{r}} \mathcal{F}^2(\Omega,r;S)-1}},\\\\
        %%%%%%%%%%%%%%%%%%%%%%%%%%%%%%%%%%%
        p^t&=-\frac{\mu  r^2 e^{\frac{2 b_0}{r}} \mathcal{F}(\Omega,r;S)}{\sqrt{r^4 e^\frac{2 b_0}{r} \mathcal{F}^2(\Omega,r;S)-1}},
        \end{aligned}
    \end{equation}
    
    where we defined $\mathcal{F}(\Omega,r;S)$ and $\mathcal{A}(\Omega,r;S)$ as 
    \begin{equation}
        \label{S3Ae6}
        \mathcal{F}(\Omega,r;S)=\frac{\mathcal{A}(\Omega,r;S)}{
        b_0 \left(2 b_0 \mu +S \Omega  e^{\frac{b_0}{r}} \sqrt{r \left(r-b_0\right)}-2 \mu  r\right)}
    \end{equation}
    and
    \begin{equation}
        \label{S3Ae7}
        \begin{aligned}
        \mathcal{A}(\Omega,r;S)&=\left(\frac{b_0 S e^{-\frac{b_0}{r}} \sqrt{r-b_0} \left(-7 b_0 r+2 b_0^2+4 r^2\right)}{r^{9/2}}+2 \mu  \Omega  \left(r-b_0\right)\right).
        \end{aligned}
    \end{equation}
    \end{widetext}
    These expressions were obtained using Eqs.~(19), (22), and (23) of Ref.~\cite{Timogiannis:2021ung}.

    Under the TD-SSC, the energy and the total
    angular momentum of the spinning test particle around the wormhole solution with $\gamma=1$ are   
    \begin{align}
    \begin{cases}
    & E = -p_t+ e^{-\frac{b_0}{r}}  \frac{S b_0}{\mu r^2} \sqrt{\frac{r-b_0}{r^3}}  p_\varphi, \\
    & J_z = p_\varphi-e^{-\frac{b_0}{r}}  \frac{S}{\mu}\sqrt{\frac{r-b_0}{r}}  p_t.
    \end{cases} \label{eq:ConservedQuantitiesTDSSC}
    \end{align}

    To obtain the equatorial circular orbits and the ISCO, we use the effective potential $V^{\text{TD}}_\text{eff}$. In the TD-SSC, the effective potential is defined from the following relation~\cite{Toshmatov:2019bda, Benavides-Gallego:2021lqn, Abdulxamidov:2022ofi}
    \begin{equation}
        \label{S3Ae9}
       (p_r)^2\propto(E-V_+)(E-V_-),
    \end{equation}
    where $V_\pm$ is the root of $(p_r)^2=0$. Since Eq.~\eqref{S3Ae9} is a quadratic equation in $E$, $V_\pm$ is given by
    \begin{equation}
        \label{S3Ae10}
        V_{\pm}=-\frac{a J_z}{b}\pm\sqrt{\frac{a^2 J^2_z}{b^2}+\frac{c-d J^2_z}{b}},   
    \end{equation}
    where $a$, $b$, $c$, and $d$ depend on the metric components and their derivative with respect to the radial coordinate $r$, see Eq.~(35) of Ref.~\cite{ Benavides-Gallego:2021lqn}. In the following, we shall focus on the case in which test particles have positive energy and therefore explore the effective potential given by $V^\text{TD}_\text{eff}=V_+$. Hence, the ISCO can be obtained by solving (numerically) the system of non-linear equations 
    \begin{equation}
        \label{S3Ae11}
        \begin{array}{ccc}
             \frac{dV^\text{TD}_\text{eff}}{dr}=0&\text{and}&            \frac{d^2V^\text{TD}_\text{eff}}{dr^2}=0 
        \end{array}
    \end{equation}
    for $r$ and $J_z$, for a given value of the particle's spin $S$.
    
%%%%%%%%%%%%%%%%%%%%%%%%%%%% MP SSC %%%%%%%%%%%%%%%%%%%%%%%%%%%%%%%%%%    
\subsection{Mathisson-Pirani SSC}\label{sec:3-2}
    Equation~(\ref{eq:generalSSC}) introduces the Mathisson-Pirani condition (MP-SSC) by selecting a future pointing time-like reference vector, which is precisely the four-velocity, $V^\mu = u^\mu $. This implies that the spin is defined as spatial for an observer who is moving in the same direction as the particle's four-velocity~\cite{OKS04}.

    Additionally, as it is shown in~\cite{OKS02}, the equation that provides the evolution of the four-velocity under the MP SSC is 
    \begin{equation}
     \frac{D u^\mu}{d\lambda}=-\frac{1}{S^2}\left(\frac{1}{2 m} R_{\rho \nu \kappa \sigma} S^\rho u^\nu S^{\kappa \sigma} S^\mu+p_\kappa S^{\mu \kappa}\right).\\
     \label{eq:MPSCCvelocitydot}
    \end{equation}
    $\hspace{1cm}$
    
    Following the procedure carried out in \cite{Timogiannis:2021ung}, and assuming that the MP-SSC holds, we can determine the values of the four-momentum components by using the definition of the kinematical rest mass, $m=-p^{\mu}u_{\mu}$, and Eqs.~\eqref{eq:SpinTensorComponents} and \eqref{eq:circularMPDequation2}. In the case of the solution with $\gamma=1$, this results in

    \begin{widetext}
    \centering
    \begin{equation}
    p^{t}=\frac{e^{\frac{b_0}{r}}\left[-m r^{3 / 2}\left(e^{\frac{2 b_0-1}{r}} r^2 \Omega^2\right)+e^{\frac{3 b_0}{r}} r^3 S \Omega^3 \sqrt{r-b_0}-e^{\frac{b_0}{r}} S b_0 \Omega \sqrt{r-b_0} \right]}{\left(r-e^{\frac{2 b_0}{r}} r^3 \Omega^2\right)^{3 / 2}},
    \label{MPmomentum1}
    \end{equation}
    
    \begin{equation}
    p^{\varphi}=\frac{e^{\frac{b_0}{r}} r^3 \Omega\left[m \sqrt{r}\left(1-e^{\frac{2 b_0}{r}} r^2 \Omega^2\right)+e^{\frac{b_0}{r}} S \Omega \sqrt{r-b_0}\right]-S b_0 \sqrt{r-b_0} }{r^{7 / 2}\left(1-e^{\frac{2 b_0}{r}} r^2 \Omega^2\right)^{3 / 2}},
    \label{MPmomentum2}
    \end{equation}
    \end{widetext}
    where the contributions of $p^{t}_{\text{hidden}}$ and $p^{\varphi}_{\text{hidden}}$ have already been added. Then, if we replace the two previous expressions in Eq.~\eqref{eq:circularMPDequation1}, we can obtain the following quartic equation for the orbital frequency
    \begin{equation}
    \xi_0+\xi_1 \Omega+\xi_2 \Omega^2+\xi_3 \Omega^3+\xi_4 \Omega^4=0, \label{eq:MPOmegaPoly} 
    \end{equation}
    with
    \begin{equation}
    \begin{aligned}
    \xi_4=&2 e^{\frac{4 b_0}{r}} m r^5, \\
     \xi_3=&-e^{\frac{3 b_0}{r}} S \sqrt{r\left(r-b_0\right)}\left(2 r^2-7 r b_0+2 b_0^2\right),  \\
      \xi_2=&-2 e^{\frac{2 b_0}{r}} m r^2\left(r+b_0\right), \\
       \xi_1=&-3 e^{\frac{b_0}{r}} S b_0 \sqrt{1-\frac{b_0}{r}},  \\
        \xi_0=&2 m b_0.  \\
    \end{aligned}
    \end{equation}
    Out of the four roots of the orbital frequency polynomial, only two are physically meaningful. These correspond to the corotation frequency $\Omega_+$ and the counterrotation frequency $\Omega_-$. Fortunately, it is possible to obtain these roots for wormhole solutions with $\gamma=1$ and $\gamma=2$ analytically. Nevertheless, we do not present them here due to their extensive form. 

    In Appendix \ref{sec:appendixMP}, we show the analogous expressions of $p^{t}$, $p^{\varphi}$ and the polynomial coefficients of the orbital frequency  for the case of the wormhole solution with $\gamma=2$.

    On the other hand, under the MP-SSC, the energy and z component of the total angular momentum of the spinning test particle around the wormhole solution with $\gamma=1$ are  
    \begin{align}
    \begin{cases}
    & E = -p_t+ e^{-\frac{b_0}{r}}  \frac{S b_0}{r^2} \sqrt{\frac{r-b_0}{r^3}}  u_\varphi, \\
    & J_z = p_\varphi-e^{-\frac{b_0}{r}}  S \sqrt{\frac{r-b_0}{r}}  u_t,
    \end{cases} \label{eq:ConservedQuantitiesMPSSC}
    \end{align}
    where $p_t=g_{tt}p^{t}$ and $p_\varphi=g_{\varphi \varphi}p^{\varphi}$ can be calculated using Eqs.~\eqref{MPmomentum1} and \eqref{MPmomentum2}, respectively. To obtain equatorial circular orbits and determine the ISCO properties, we employ the treatment presented in~\cite{Harms:2016ctx}. This treatment, when applied to the MP-SSC, involves the use of three effective potentials, in contrast to the TD-SSC where only one is needed. The first potential is derived from the radial component of Eq.~\eqref{eq:MPSCCvelocitydot}, which yields $\frac{d u^r}{d \lambda}=-\frac{V_{\text{eff}}^{\text{MP1}}}{2 S g_{r r} \sqrt{-g}}$ where
    \begin{align}
    V_{\text{eff}}^{\text{MP1}}:= &2 g_{r r} \sqrt{g_{\theta \theta}}\left(g_{\varphi \varphi} u^\varphi p_t-g_{t t} u^t p_\varphi\right)\\
    &-S \sqrt{-g}\left[\frac{\partial g_{t t}}{\partial r} \left(u^{t}\right)^2+\frac{\partial g_{\varphi \varphi}}{\partial r} \left(u^{\varphi}\right)^2\right].
    \label{MPpotential1}
    \end{align}
    By rewriting the kinematical rest mass definition, the second potential can be obtained as
    \begin{equation}
    V_{\text{eff}}^{\text{MP2}}:= u^rp_r=-(m+u^tp_t+u^\varphi p_\varphi).
    \label{MPpotential2}
    \end{equation}
    In the last expressions for the potential, $p_t$ and $p_\varphi$ are replaced using Eqs.~\eqref{MPmomentum1} and \eqref{MPmomentum2} to get $V_{\text{eff}}=V_{\text{eff}}(r,E,J_z,S,u_t,u_\varphi)$. 

    Finally, we can determine the third potential by utilizing the normalization condition of the reference vector with the MP-SCC, which matches with the four-velocity normalization, $u_\alpha u^\alpha = -1$. From here we have that $
    u^r= \pm \sqrt{\frac{V_{\text{eff}}^{\text{MP3}}}{g_{r r}}}
    $
    where
    \begin{equation}
    V_{\text{eff}}^{\text{MP3}}:=-\left[g_{t t} \left(u^{t}\right)^2+g_{\varphi \varphi} \left(u^{\varphi}\right)^2+1\right].
    \label{MPpotential3}
    \end{equation}
    Hence, to determine the ISCO properties, we solve a system of nine equations derived from the three previous potentials equaled to zero, along with their first and second derivatives with respect to $r$ also equaled to zero. Then, for a given spin $S$, this calculation yields the values of nine unknown variables at the ISCO, namely $r$, $E$, $J_z$, $u_t$, $u_t'$, $u_t''$, $u_\varphi$, $u_\varphi'$ and $u_\varphi''$, where prime denotes the derivation with respect to $r$.

%%%%%%%%%%%%%%%%%%%%%%%%% OKS SSC %%%%%%%%%%%%%%%%%%%%%%%%%%%%%%%%
\subsection{Ohashi-Kyrian-Semerák SSC} \label{sec:3-3}

    The Ohashi-Kyrian-Semerak condition (OKS-SSC) is introduced through Eq.~(\ref{eq:generalSSC}) by choosing a future pointing time-like reference vector, $V^\mu = w^\mu$, satisfying the conditions \cite{OKS01, OKS02, OKS03, OKS04}
    \begin{align}
    w_\mu w^\mu = &-1 \label{eq:OKS_normalization}\\
    \frac{D w^\mu}{d\lambda} = & 0.\label{eq:OKS_condition}
    \end{align}

    Due to these assumptions, it is straightforward to show that the hidden momentum vanishes so that the momentum and the velocity are proportional \cite{Harms:2016ctx},
    \begin{equation}
    p^\mu = m u ^\mu, \label{eq:OKSmomentumvelocity}
    \end{equation}
    which implies that $\mu = m$ is a constant of motion as well as the spin angular momentum, $S$. The OKS-SSC also implies that the spin tensor Eq.~(\ref{eq:MPD2}) reduces to
    \begin{equation}
    \frac{D S^{\mu \nu}}{d\lambda} = 0.
    \end{equation}
    
    Since the components of the reference vector are not completely constrained by the definition given in (\ref{eq:OKS_normalization}) and (\ref{eq:OKS_condition}), we can choose $w^r = w^\theta = 0$; a natural choice for equatorial circular orbits. Hence, from the normalization condition (\ref{eq:OKS_normalization}) it is possible to relate the non-zero components of the reference vector as
    \begin{align}
    w^t = & \sqrt{-\frac{1+g_{\varphi \varphi} (w^\varphi)^2}{g_{tt}}}, \label{eq:referenceVectorOKS1} 
    \end{align} 
    while the second MPD Eq.~(\ref{eq:circularMPDequation2}) gives the component $w^\varphi$ in terms of the orbital frequency of the test particle, $\Omega$, as \cite{Harms:2016ctx} 
    \begin{align}
    w^\varphi = &\pm \sqrt{-\frac{g_{tt} (\Gamma^t_{tr} + \Omega \Gamma^t_{\varphi r} )^2}{ g_{\varphi \varphi}^2(\Gamma^\varphi_{tr} + \Omega \Gamma^\varphi_{\varphi r} )^2 + g_{tt}g_{\varphi \varphi} (\Gamma^t_{tr} + \Omega \Gamma^t_{\varphi r} )^2 }}.\label{eq:referenceVectorOKS2}
    \end{align} 

    In Appendix~\ref{sec:appendix}, we obtain a general expression for the 6th order polynomial (\ref{eq:OKSOmegaPoly}) which gives the orbital frequency under the OKS-SSC. Once the geometric expressions, such as connections and Riemann tensor, for the metric (\ref{eq:WHmetric}) are replaced, we obtain a very long expression that is not highly elucidating and therefore, it will not be provided in this paper (but is available in the supplementary material). Finally, the energy and the angular momentum for the spinning test particle under the OKS-SSC are given, in terms of the reference vector components (\ref{eq:referenceVectorOKS1}) and (\ref{eq:referenceVectorOKS2}), as
    \begin{align}
    \begin{cases}
    & E = -p_t+ e^{-\frac{b_0}{r}}  \frac{S b_0}{r^2} \sqrt{\frac{r-b_0}{r^3}}  w_\varphi, \\
    & J_z = p_\varphi-e^{-\frac{b_0}{r}}  S \sqrt{\frac{r-b_0}{r}}  w_t,
    \end{cases} \label{eq:ConservedQuantitiesOKSSSC}
    \end{align}

    In order to describe equatorial circular orbits and in particular to obtain the ISCO, again, we follow the treatment presented in \cite{Harms:2016ctx}, which in this case is also based on the use of three effective potentials (similarly to the MP-SSC). The first potential is obtained from the normalization of the momentum, which gives $p_r = \pm \sqrt{\frac{V_{\text{eff}}^{\text{OKS1}} }{g^{rr}}}$ where
    \begin{equation}
    V_{\text{eff}}^{\text{OKS1}}  := - \left( m^2 + g^{tt} p_t^2 + g^{\varphi \varphi} p_\varphi^2 \right) = 0 \label{eq:OKSpotential1}
    \end{equation}
    and we demand to be zero in order to represent circular trajectories.  

    The second potential arises from the normalization condition for the reference vector of Eq.~ (\ref{eq:OKS_normalization}), establishing
    \begin{equation}
        V_{\text{eff}}^{\text{OKS2}}  := 1 + g^{tt} w_t^2 + g^{\varphi \varphi} w_\varphi^2 = 0. \label{eq:OKSpotential2}
    \end{equation}
    The third potential is obtained from the radial component of the Eq.~ (\ref{eq:OKS_condition}), 
    \begin{equation}
        \frac{dw_r}{d\lambda}  = -\frac{V_{\text{eff}}^{\text{OKS3}} }{2m g_{tt}^2 g_{\varphi\varphi}^2} = 0,
    \end{equation}
    where
    \begin{equation}
        V_{\text{eff}}^{\text{OKS3}}  :=  w_t g_{\varphi \varphi}^2 p_t \partial_r g_{tt}  + w_\varphi g_{tt}^2 p_\varphi \partial_r g_{\varphi\varphi}  = 0. \label{eq:OKSpotential3}
    \end{equation}
    
    The ISCO properties are obtained by solving the set of nine equations given by the three potentials and its first and second derivatives with respect to $r$ equal to zero. This system will give the nine unknown variables $r$, $E$, $J_z$, $w_t$, $w_t'$, $w_t''$, $w_\varphi$, $w_\varphi'$ and $w_\varphi''$ at the ISCO.
    
%%%%%%%%%%%%%%%%%%%%%%%%% Tables %%%%%%%%%%%%%%%%%%%%%%%%%%%%%%%%%
    \begin{center}
    \begin{table*}
    \begin{tabular}{||c c c c||} 
     \hline
     $\hat{\Omega}_n$ & TD  & MP  & OKS \\ [0.5ex]  
     \hline\hline \rule{0pt}{3ex}  
     $ \mathcal{O} \left( \sigma^0 \right) $ & $\frac{1}{\bar{r}^{3/2}} $ & $\frac{1}{\bar{r}^{3/2}} $ & $\frac{1}{\bar{r}^{3/2}} $ \\ [1.5ex]
     \hline \rule{0pt}{3ex}  
      $\mathcal{O} \left( \sigma^1 \right) $ & $-\frac{\left(5 \bar{r}-2\right) \sqrt{\bar{r}-1}}{4 \sqrt{\bar{r}^9}} $ & $-\frac{\left(5 \bar{r}-2\right)\sqrt{\bar{r}-1}}{4 \sqrt{\bar{r}^9} }$ & $-\frac{\left(5 \bar{r}-2\right)\sqrt{\bar{r}-1} }{4 \sqrt{\bar{r}^9} }$ \\ [1.5ex]
     \hline  \rule{0pt}{4ex}  
      $\mathcal{O} \left( \sigma^2 \right) $ & $\frac{ \left(\bar{r}-1\right) \left(\bar{r} \left(65 \bar{r}-36\right)+4\right)}{32 \bar{r}^{15/2}}$ & $\frac{ \left(\bar{r}-1\right) \left(\bar{r} \left(65 \bar{r}-36\right)+4\right)}{32 \bar{r}^{15/2}}$ & $\frac{ \left(5 \bar{r}-2\right) \left(11 \bar{r}^2-\bar{r}-2\right)}{32 \bar{r}^{15/2}}$ \\ [1.5ex]
     \hline  \rule{0pt}{4ex}  
      $\mathcal{O} \left( \sigma^3 \right) $ & $\hspace{0.5cm}-\frac{\sqrt{\bar{r}-1}\left(5 \bar{r}-2\right) }{8 \sqrt{\bar{r}^{15}}} \hspace{0.5cm}$ & $\hspace{0.3cm}-\frac{\sqrt{\bar{r}-1} \left(5 \bar{r}-2\right) \left(7 \bar{r}-4\right)}{8 \sqrt{\bar{r}^{17}}}\hspace{0.3cm}$ & $\hspace{0.3cm}-\frac{ \left(5\bar{r}-2\right) \left(32 \bar{r}^3-17 \bar{r}^2-8 \bar{r}+4\right)}{32  \sqrt{\bar{r}^{19}}\sqrt{\bar{r}-1}}\hspace{0.3cm}$  \\ [1.5ex] 
     \hline 
     \hline
    \end{tabular}
    \caption{Orbital frequency orders in the expansion in powers of $\sigma$ for arbitrary circular equatorial orbits for the case $\gamma = 1$}
    \label{table:Frequency01}
    \end{table*}
    \end{center}

    \begin{center}
    \begin{table*}
    \begin{tabular}{||c c c c||} 
     \hline
     $\hat{\Omega}_n$ & TD  & MP  & OKS \\ [0.5ex]  
     \hline\hline \rule{0pt}{3ex}  
     $ \mathcal{O} \left( \sigma^0 \right) $ & $\frac{1}{\bar{r}^{3/2}} $ & $\frac{1}{\bar{r}^{3/2}} $ & $\frac{1}{\bar{r}^{3/2}} $ \\ [1.5ex]
     \hline \rule{0pt}{3ex}  
      $\mathcal{O} \left( \sigma^1 \right) $ & $-\frac{ \left(2 \bar{r}^3-3 \bar{r}+1\right)}{2 \bar{r}^5 \sqrt{\bar{r}^2-1}} $ & $-\frac{\left(2 \bar{r}^3-3 \bar{r}+1\right)}{2 \bar{r}^5 \sqrt{\bar{r}^2-1}} $ & $-\frac{ \left(2 \bar{r}^3-3 \bar{r}+1\right)}{2 \bar{r}^5 \sqrt{\bar{r}^2-1}}$ \\ [1.5ex]
     \hline  \rule{0pt}{4ex}  
      $\mathcal{O} \left( \sigma^2 \right) $ & $\frac{ \left(\bar{r}-1\right) \left(2 \bar{r} \left(\bar{r}+1\right)-1\right) \left(6 \bar{r} \left(\bar{r}+1\right)-1\right)}{8 \bar{r}^{17/2} \left(\bar{r}+1\right)} $ & $\frac{ \left(\bar{r}-1\right) \left(2 \bar{r} \left(\bar{r}+1\right)-1\right) \left(6 \bar{r} \left(\bar{r}+1\right)-1\right)}{8 \bar{r}^{17/2} \left(\bar{r}+1\right)} $ & $\frac{ \left(-12 \bar{r}^5-20 \bar{r}^4+8 \bar{r}^2+\bar{r}-1\right)}{8 \bar{r}^{17/2} \left(\bar{r}+1\right)}$ \\ [1.5ex]
     \hline  \rule{0pt}{4ex}  
      $\mathcal{O} \left( \sigma^3 \right) $ & $\hspace{0.5cm}-\frac{ \left(2 \bar{r}^3-3 \bar{r}+1\right)}{2 \bar{r}^9 \sqrt{\bar{r}^2-1}} \hspace{0.5cm}$ & $\hspace{0.3cm}-\frac{ \left(\bar{r}-1\right) \left(2 \bar{r}^2+2 \bar{r}-1\right) \left(3 \bar{r}^2+2 \bar{r}-2\right)}{2 \bar{r}^{10} \sqrt{\bar{r}^2-1}} \hspace{0.3cm}$ & $\hspace{0.3cm}-\frac{ \left(2 \bar{r} \left(\bar{r}+1\right)-1\right) \left(2 \left(\bar{r} \left(4 \bar{r}^2+6 \bar{r}-1\right)-4\right) \bar{r}^2+1\right)}{4 \bar{r}^{11} \sqrt{\left(\bar{r}^2-1\right) }\left(\bar{r}+1\right)} \hspace{0.3cm}$  \\ [1.5ex] 
     \hline 
     \hline
    \end{tabular}
    \caption{Orbital frequency orders in the expansion in powers of $\sigma$ for arbitrary circular equatorial orbits for the case $\gamma = 2$}
    \label{table:Frequency02}
    \end{table*}
    \end{center}
    	
%%%%%%%%%%%%%%%%%%%%%%%%%%%%%%%%%%%%%%%%%%%%%%%%%%%%%%%%%%%%%%%%%%%%%%%%%%%%%%%%%%%%%%%% Section IV %%%%%%%%%%%%%%%%%%%%%%%%%%%%%%%%%%
%%%%%%%%%%%%%%%%%%%%%%%%%%%%%%%%%%%%%%%%%%%%%%%%%%%%%%%%%%%%%%%%%% 

%%%%%%%%%%%%%%%%%%%%%%%%%%figures%%%%%%%%%%%%%%%%%%%%%%%%%%%%%%%%%%
    \begin{figure*}
    \centering
    \includegraphics[width=1\linewidth]{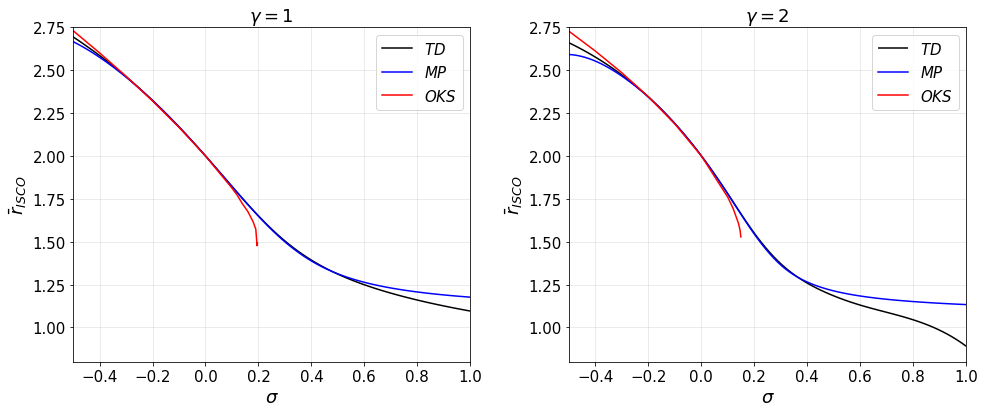}
    \caption{ISCO radius for different values of spin $\sigma$ for both cases $\gamma=1,2$.}
    \label{fig:fig1}
    \end{figure*}
    
    \begin{figure*}
        \centering
    \includegraphics[width=1\linewidth]{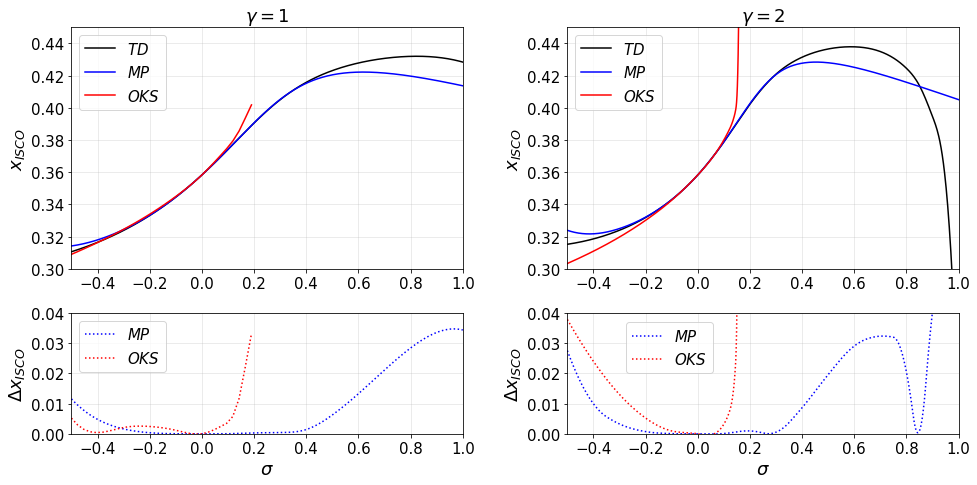}
        \caption{On top panels, we present the ISCO orbital frequency  parameter $x_{ISCO}$ for each SSC as reference. While in the lower panels, we show the relative differences $\Delta x_{ISCO}$ of MP and OKS SSCs with respect to TD SSC, using the Eq.~\eqref{eq:relativediff1}.  These quantities are given as function of spin $\sigma$ and for both cases $\gamma=1,2$.}
        \label{fig:fig2}
    \end{figure*}
%%%%%%%%%%%%%%%%%%%%%%%%%%%%%%%%%%%%%%%%%%%%%%%%%%%%%%%%%%%%%%%%%% 
    \section{ORBITAL PARAMETERS USING THE SPIN SUPPLEMENTARY CONDITIONS} \label{sec:4}
    
    \subsection{Orbital Frequency} \label{sec:4.1}

    As was presented above, each of the SSC gives a polynomial for the orbital frequency. For the TD-SSC, the 2nd-degree polynomial is given by Eq.~(\ref{eq:TDOmegapoly}), for the MP-SSC, we obtain the 4th-degree polynomial (\ref{eq:MPOmegaPoly}) and in the case of the OKS-SSC, Eq.~(\ref{eq:OKSOmegaPoly}) is a 6th-degree polynomial. Following the method presented in \cite{{Timogiannis:2021ung}}, we introduce the dimensionless quantities $\bar{r} = \frac{r}{b_0}$, $\hat{\Omega} = b_0 \Omega$ and $\sigma = \frac{s}{m b_0}$ (for the MP-SSC) or $\sigma = \frac{s}{\mu b_0}$ (for the TD-SSC and the OKS-SSC) in these equations. Then, we expand the resulting polynomials in powers of the dimensionless spin $\sigma$ by introducing $\hat{\Omega} = e^{-\frac{1}{r}}\hat{\Omega}_n \sigma^n + \mathcal{O} (\sigma^4)$ with $n=0,1,2,3$ (we only consider terms up to the 3rd order in $\sigma$ because there we obtain the differences between the results arising from the three SSCs).

    It is clear that, depending on the order of the polynomial, we obtain many roots and we need to choose between them which one is physically relevant. In particular, the selection criterion for the order $\hat{\Omega}_0$ will be that the Keplerian frequency is recovered for vanishing spin and for the first order of approximation in the exponential terms (involving the approximation $r \gg b_0$).
    
    From the results compiled in Table \ref{table:Frequency02} it is clear that the three SSCs produce equivalent results up to the linear order $\mathcal{O}(\sigma^1)$. At the order $\mathcal{O}(\sigma^2)$, TD- and MP-SSCs give the same contribution to the frequency but the OKS-SSC gives a different result, and when calculating the contribution of order   $\mathcal{O}(\sigma^3)$ all three results  differ.

    \subsection{ISCO parameters} \label{sec:4.2}

    The ISCO radius, $\bar{r}_{ISCO}$, is an essential parameter that characterizes the gravitational dynamics near compact objects like wormholes. This quantity plays a crucial role in various possible astrophysical phenomena associated with wormholes, such as the accretion of matter, gravitational lensing, gravitational radiation, and the formation of jets, among others. Thus, the study of $\bar{r}_{ISCO}$ provides valuable insights into the properties of the wormhole. This is why it is chosen as the first parameter to compare the different SSCs.

    Recall that to find $\bar{r}_{ISCO}$ we have to use a different approach depending on each SSC. For the TD-SSC, we will use the effective potential of Eq.~\eqref{S3Ae10} together with the conditions of Eqs.~\eqref{S3Ae11}. For the MP-SSC, the three potentials of the Eqs.~\eqref{MPpotential1}, \eqref{MPpotential2} and \eqref{MPpotential3} are used, together with their first and second derivatives 
    equal to zero. And finally, for the OKS-SSC, the three potentials of the Eqs.~\eqref{eq:OKSpotential1}, \eqref{eq:OKSpotential2} and \eqref{eq:OKSpotential3} will be required, along with their first and second derivatives equal to zero too. 

    In Fig.~\ref{fig:fig1}, we share the ISCO radius $\bar{r}_{ISCO}$ as a function of the particle's spin $\sigma$ for each SSC and both wormhole solution cases $\gamma=1,2$; as a first result, we can see that the value of $\bar{r}_{ISCO}$ calculated with the three different SSCs behaves very similarly, especially in the vicinity of $\sigma=0$. In all cases, it is possible to see clearly that $\bar{r}_{ISCO}$ decreases as the value of spin $\sigma$ increases. Moreover, although the figure shows that $\bar{r}_{ISCO}$ behaves similarly for $\gamma=1$ and $\gamma=2$, the former generally has slightly larger values. On the other hand, for high values of the spin, note that the MP- and TD-SSCs behave similarly in contrast to the OKS-SSC. However, for the MP-SSC, the numerical computation of $\bar{r}_{ISCO}$ fails with $\sigma<-0.6$ and $\sigma<-0.5$ for the two wormhole solutions, $\gamma=1$ and $\gamma=2$, respectively. Meanwhile, in the case of the OKS-SSC, the numerical computation of $\bar{r}_{ISCO}$ fails when $\sigma>0.2$ and $\sigma>0.15$ for $\gamma=1$ and $\gamma=2$, respectively. The failures in the numerical calculations for the MP- and OKS-SSCs are similar to those obtained for the ISCO of spinning particles around Kerr black holes in \cite{Lukes2017}. Indeed, the reason for the different behavior among the SSCs lies in the fact that each condition represents a different reference point for the position of the centroid. These differences stem from attempting to describe extended bodies using only their first two multipoles. However, it is important to remark that extended bodies have an infinite number of multipoles that are neglected intentionally in the pole-dipole approximation used to obtain the MPD eqiuations~\cite{Lukes2017}.
    
    A second way to compare the different SSCs is by using a procedure similar to the one carried out in \cite{Lukes2017}. There, the authors argue that to provide a full gauge invariant discussion one can use the following ISCO orbital parameter
    \begin{equation}
      x_{ISCO} \equiv(b_0 \hat{\Omega}_{ISCO})^{2 / 3}.
    \end{equation}
    Then, the relative difference of the ISCO frequency parameters given by the MP- or OKS-SSCs with respect to the ISCO frequency parameters given by the TD-SSC is defined by
    \begin{equation}
    \Delta x_{ISCO}=\frac{\left|x_{ISCO}^{SSC}-x_{ISCO}^{TD}\right|}{x_{ISCO}^{TD}},
    \label{eq:relativediff1}
    \end{equation}
    where $x_{ISCO}^{SSC}$ can correspond to the ISCO frequency parameter given by the MP- or OKS-SSCs. 

    In Fig.~ \ref{fig:fig2}, top panels, we show the ISCO orbital frequency parameter $x_{ISCO}$ for each SSC. In the lower panels, the figure also shows the relative differences $\Delta x_{ISCO}$ of MP- and OKS-SSCs with respect to the TD-SSC. The results are plotted as a function of the particle's spin $\sigma$ and for both wormhole solution cases $\gamma=1,2$. Since $x_{ISCO}=x(\bar{r}_{ISCO})$, the behavior of the orbital frequency parameter is similar to that of $\bar{r}_{ISCO}$ for each SSC. Therefore, the numerical computations for finding $x_{ISCO}$ have the same validity regions as those for $\bar{r}_{ISCO}$. Additionally, we observe that the parameter $x_{ISCO}$ increases until reaching a maximum point, especially for TD- and MP-SSCs, which coincides with the location of the inflection points found in $\bar{r}_{ISCO}$. This maximum point is greater for the TD-SSC. Notably, the TD- and OKS-SSCs exhibit a divergence at $\sigma \approx 0.15$ and $\sigma \approx 1$, respectively, in the case of the solution with $\gamma=2$. When examining the relative differences near $\sigma=0$, the differences between the MP- and TD-SSCs are smaller than those between the OKS- and TD-SSCs. In any case, the differences between the SSCs become more significant in the regions close to the inflection points of $\bar{r}_{ISCO}$.

    Each SSC is defined using a different reference centroid. Consequently, to obtain equivalent results from the different SSCs, it is necessary to apply a correction to the centroid. Nevertheless, although centroid corrections give account for the differences between SSCs, it is worth noting that corrections could deviate a circular trajectory from circularity; this is because both the spin and the position of the centroid would undergo changes, and the worldline that was previously on the ISCO for one centroid may not remain on the ISCO for another centroid \cite{Lukes2017}. Therefore, when approximating the results of each SSC, it's crucial to apply a correction to both the spin and the position of the centroid, as long as the pole-dipole approximation allows it \cite{Timogiannis:2021ung}.
      
%%%%%%%%%%%%%%%%%%%%%%%%%%%%%%%%%%%%%%%%%%%%%%%%%%%%%%%%%%%%%%%%%%%%%%%%%%%%%%%%%%%%%%%%%% Section V %%%%%%%%%%%%%%%%%%%%%%%%%%%%%%%%
%%%%%%%%%%%%%%%%%%%%%%%%%%%%%%%%%%%%%%%%%%%%%%%%%%%%%%%%%%%%%%%%%  
    \begin{figure*}
        \centering
    \includegraphics[width=1\linewidth]{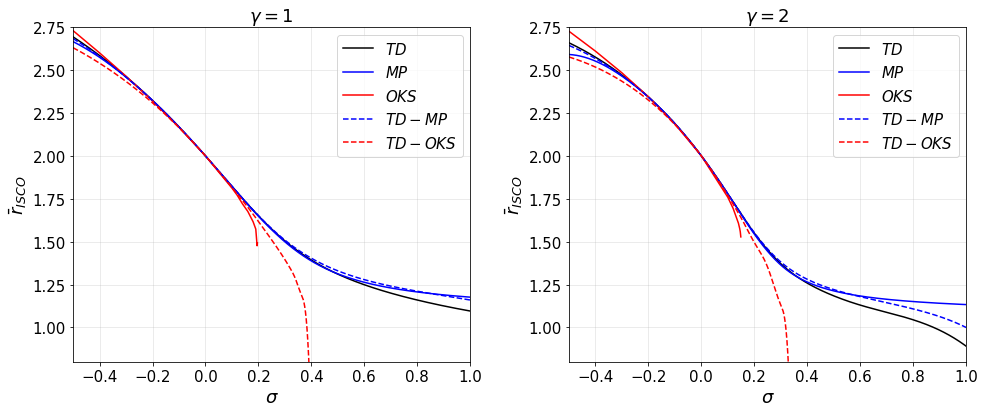}
        \caption{ISCO radius for each SSCs compared to those corrected by $\tilde{r} \neq \bar{r}$, for different values of spin $\sigma$ and for both cases $\gamma=1,2$.}
        \label{fig:fig33}
    \end{figure*}
    
    \begin{figure*}
        \centering
    \includegraphics[width=1\linewidth]{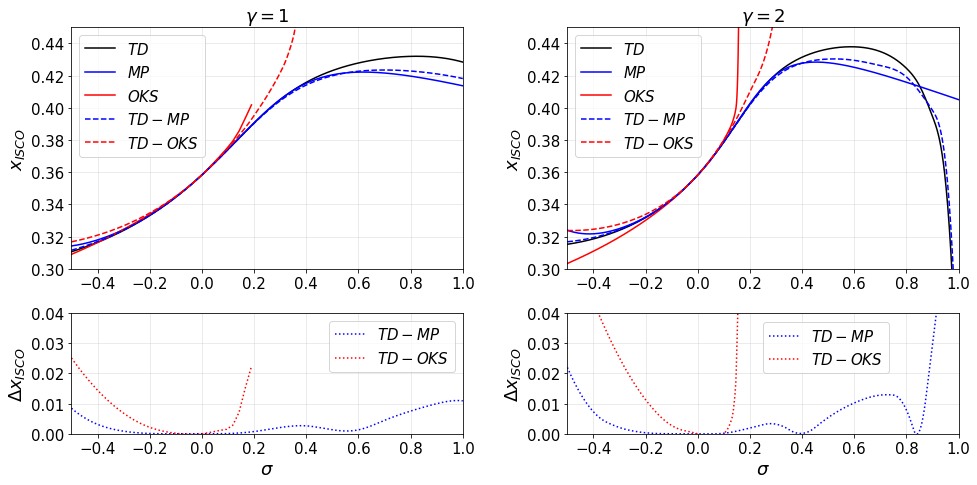}
        \caption{On top panels, we present the ISCO orbital frequency  parameter $x_{ISCO}$ for each SSCs compared to those  corrected by $\tilde{r} \neq \bar{r}$. While in the lower panels, we show the relative differences $\Delta x_{ISCO}$ of the MP and OKS SSCs with respect to the TD-MP and TD-OKS SSCs corrections, using the Eqs.~\eqref{eq:relativediff2} and \eqref{eq:relativediff3}, respectively.  These quantities are given as a function of spin $\sigma$ and for both cases $\gamma=1,2$.}
        \label{fig:fig44}
    \end{figure*}
    
    \section{CENTROIDS’ CORRECTIONS} \label{sec:5}

    The differences in the orbital frequency obtained using the three spin supplementary conditions, shown in Tables \ref{table:Frequency01} and \ref{table:Frequency02}, are usually explained by noting that each SSC defines a particular centroid \cite{Timogiannis:2022bks, Lukes2017} and therefore, all the moments are evaluated correspondingly. We will assume that the choice of the SSC corresponds to a centroid's correction in the form $z^\mu \rightarrow \tilde{z}^\mu = z^\mu + \delta z^\mu$. In order to compare the results obtained using the three SSCs presented in this paper, we will consider the TD-SSC as a reference because the corresponding centroid is uniquely determined. Hence, the quantities under the TD-SSC will be denoted by a tilde, $\tilde{}$. 

    Although the momentum components do not depend on the choice of the centroid, the spin tensor does change according to
    \begin{equation}
    S^{\mu \nu} \rightarrow \tilde{S}^{\mu \nu} = S^{\mu \nu} + p^\mu \delta z^\nu - p^\nu \delta z^\mu.
    \label{eq54}
    \end{equation}
    Hence, following the discussion in the appendix B of \cite{Timogiannis:2021ung}, we will impose the constraint $V_\alpha \delta^\alpha = 0$, which is equivalent to the relation 
    \begin{equation}
    \tilde{p}_\alpha \delta z ^\alpha = 0    
    \end{equation}
    together with the condition that the centroid cannot have non-radial shifts between the TD- and the MP-/OKS-SSCs. This implies that the change in the centroid will be
    \begin{equation}
    \label{s5e3}
    \delta z^\alpha = \delta r = \frac{\tilde{p}_\mu S^{\mu r}}{\tilde{\mu}^2},
    \end{equation}
    where $\tilde{\mu}^2 = - \tilde{g}_{\alpha \beta} p^\alpha p ^\beta$. Furthermore, in the previous section, we analyzed the relative differences of the ISCO frequency parameters $\Delta x_{ISCO}$ for the MP- and OKS-SSCs with respect to the TD-SSC, as done in \cite{Lukes2017}. Meanwhile, in this section, to compare the different centroid corrections, we will not calculate the relative difference, $\Delta x_{ISCO}$, with respect to the TD-SSC but in terms of each SSC correction. In this sense, to relate the ISCO frequency parameters $x_{ISCO}$ given by the MP-SSC centroid correction applied to the TD-SSC (denoted as TD-MP) with $x_{ISCO}$ given by the MP-SSC without correction, we introduce the relative difference:
    
    \begin{equation}
    \Delta x_{ISCO}=\frac{\left|x_{ISCO}^{MP}-x_{ISCO}^{TD-MP}\right|}{x_{ISCO}^{TD-MP}}.
    \label{eq:relativediff2}
    \end{equation}
    Similarly, the relative differences given by the OKS-SSC without any correction with respect to the OKS-SSC centroid correction applied to the TD-SSC (denoted as TD-OKS) are
    \begin{equation}
    \Delta x_{ISCO}=\frac{\left|x_{ISCO}^{OKS}-x_{ISCO}^{TD-OKS}\right|}{x_{ISCO}^{TD-OKS}}.
    \label{eq:relativediff3}
    \end{equation}
    
    Additionally, we need to keep in mind that in the MP-SSC, the dimensionless spin is expressed as $\sigma=\frac{S}{m b_0}$, while in the TD-SSC, it takes the form of $\sigma=\frac{S}{\mu b_0}$. To obtain the dimensionless spin $\tilde{\sigma}$ measured in the TD reference frame in terms of $\sigma$ measured in the MP frame, we need to use the relationship between $m$ and $\mu$, which is given by~\cite{Costa:2017kdr, Timogiannis:2021ung}
    \begin{equation}
    \mu^2=m^2+\frac{S^{\alpha \lambda} S_{\lambda \beta} p^\beta p_\alpha}{S^2}.
    \label{eq:muandm}
    \end{equation}
    The last expression is very useful for our numerical analysis of the centroid’s corrections to the radius and the orbital frequency of the ISCO orbit that we will present below.
    
    \subsection{Corrections to the position of the centroid} \label{sec:5.1}

    In this subsection, we will consider the corrections to the reference position of the centroid applied to the radius of the ISCO ($\bar{r}_{ISCO}$) and the corresponding orbital frequency parameter $x_{ISCO}$. To begin with, restricting to a linear radial correction of the centroid, the ISCO radius will change as
    \begin{equation}
    \tilde{r}_{ISCO} = \bar{r}_{ISCO} + \delta \bar{r}_{ISCO},
    \end{equation} 
    where Eqs.~(\ref{eq:SpinTensorComponents}) give the quantity
    \begin{align}
    \delta r = &\frac{p_t S^{tr} + p_\varphi S^{\varphi r}}{\mu ^2}\\
    =& -\frac{S}{\mu^2} \sqrt{-\frac{g_{\theta \theta}}{g}}\left[ p_t  V_{\varphi} - p_\varphi V_t \right].
    \end{align}

    The evaluation of this correction for the ISCO depends on the SSC. In the case of the MP-SSC, for example, we use the velocity as the reference vector and Eqs.~(\ref{MPmomentum1}) and (\ref{MPmomentum2}) for the momentum. Meanwhile, for the OKS-SSC we use the reference vector given in Eqs.~(\ref{eq:referenceVectorOKS1}) and (\ref{eq:referenceVectorOKS2}). 

    Figure~\ref{fig:fig33} shows the ISCO radius $\bar{r}_{ISCO}$ calculated for each SSC and $\bar{r}_{ISCO}$ derived by making the centroid correction due to the radial shift induced by the MP- and the OKS-SSCs on the TD-SSC. Meanwhile, in Fig.~ \ref{fig:fig44}, we share the ISCO orbital frequency parameter $x_{ISCO}$ for each SSC compared to those corrected by $\tilde{r} \neq \bar{r}$ and the relative differences $\Delta x_{ISCO}$ of the MP- and OKS-SSCs with respect to the TD-MP- and TD-OKS-SSCs corrections, using the Eqs.~\eqref{eq:relativediff2} and \eqref{eq:relativediff3}, respectively. The plots show that the first-order corrections to the ISCO parameters, $\bar{r}$ and $x_{ISCO}$, based on $\tilde{r}_{ISCO}$, effectively bring the TD-SSC results closer to the behavior of the other SSCs. Comparing the relative differences $\Delta x_{ISCO}$ obtained in the previous section with those obtained by applying the correction for the radial position of the centroid confirms that the results of TD-MP and TD-OKS SSCs exhibit behavior closer to that of the MP- and OKS-SSCs, in contrast to the TD-SSC. However, it's important to note that this behavior does not occur for all spin values. For instance, the correction represented by the TD-OKS-SSC for values of $\sigma<0$ exhibits a behavior that differs significantly from the OKS-SSC, compared to that obtained with the TD-SSC, which already behaves more similarly to the OKS-SSC in any case. Furthermore, the correction provided by $\tilde{r} \neq \bar{r}$ enables the calculation of ISCO parameters for values of $\sigma$ that were not previously accessible in the numerical calculations using the MP- and OKS-SSCs. For example, in the TD-OKS-SSC, $\bar{r}_{ISCO}$ and $x_{ISCO}$ could be computed for values greater than $\sigma>0.2$ and $\sigma>0.15$ for solutions with $\gamma=1$ and $\gamma=2$, respectively, where the computational routine had previously failed.

    We attempted to calculate the corrections to the second-order of $\delta r^2$ due to the radial shift of the position of the centroid. However, we did not achieve satisfactory results and opt not to share them, mainly because the second-order corrections were further away from the first-order corrections of $\delta r$; this conclusion is consistent with similar studies on black holes by Schwarzschild and Kerr \cite{Timogiannis:2021ung, Timogiannis:2022bks}, where they noted that higher order corrections did not improve the approximation between the behavior of different SSCs, since the pole-dipole approximation becomes invalid and is necessary to consider multiples of higher order in the calculations. 

    \subsection{Corrections to the Spin}
    \label{sec:5.2}

    Equation~\eqref{eq54} describes how the spin tensor changes when the centroid is measured relative to another four-vector. Because $\tilde{S} \neq S$, it is possible to derive an equation that relates the change in the measured spin value to the radial shift. We obtain this expression by applying Eq.~\eqref{eq54} to the non-vanishing components of the spin tensor and assuming that both centroids move on circular equatorial orbits, as described in \cite{Timogiannis:2021ung}. Then, if we expand the definition of the spin angular momentum of Eq.~ \ref{eq14} in terms of the radial shift, this yields~\cite{Timogiannis:2021ung}
    \begin{equation}
    \begin{aligned}
    \tilde{S}^2= & S^2+\delta r\left\{g _ { r r } \Big[ \partial_r g_{\varphi \varphi}\left(S^{r \varphi}\right)^2+\partial_r g_{t t}\left(S^{t r}\right)^2\right.  \\
    & \left.+2\left(p_t S^{t r}-p_\varphi S^{r \varphi}\right)\Big]+\frac{S^2 \partial_r g_{r r}}{g_{r r}}\right\}+\mathcal{O}\left(\delta r^2\right) .
    \label{eq:spincorrection}
    \end{aligned} 
    \end{equation}
    Next, depending on the SSC, we need to use the expressions for the spin tensor and the radial displacement.

    \subsubsection{Spin Corrections for TD-MP SCC}

    To determine the dimensionless spin $\tilde{\sigma}$ for the spin transition from the MP- to the TD-SSC, we must divide both sides of the Eq.~\eqref{eq:spincorrection} by $\tilde{\mu}^2 b_0^2$. So, expanding $1/\tilde{\mu}^2 b_0^2$ in the linear approximation of $\delta r$, and then replaced along with $\mu$ in terms of $m$ using the Eq.~\eqref{eq:muandm}, one obtains~\cite{Timogiannis:2021ung}
    \begin{widetext}
    \centering
    \begin{align}
    \label{spin_correction_TD_MP}
    \tilde{\sigma}^2= & \frac{\sigma^2}{\sigma^2-g_{r r}\left(p_t \sigma^{t r}-p_\varphi \sigma^{r \varphi}\right)^2 / m^2}\left\{\sigma^2+\delta r\left\{\frac{\sigma^4}{\sigma^2-g_{r r}\left(p_t \sigma^{t r}-p_\varphi \sigma^{r \varphi}\right)^2 / m^2}\left[\partial r g_{t t}\left(\frac{p_t}{m g_{t t}}\right)^2+\partial r g_{\varphi \varphi}\left(\frac{p_\varphi}{m g_{\varphi \varphi}}\right)^2\right]\right.\right. \notag \\
    & \left.\left.+g_{r r}\left[\partial r g_{\varphi \varphi}\left(\sigma^{r \varphi}\right)^2+\partial r g_{t t}\left(\sigma^{t r}\right)^2+\frac{2}{m b_0}\left(p_t \sigma^{t r}-p_\varphi \sigma^{r \varphi}\right)\right]+\frac{\sigma^2 \partial r g_{r r}}{g_{r r}}\right\}\right\}, 
    \end{align}
    \end{widetext}
    where  
    \begin{equation}
    \sigma^{\kappa \nu}=\frac{S^{\kappa \nu}}{m b_0} \label{eq:sigmaUpUp}    
    \end{equation}
    is the normalized spin tensor. Applying a power series expansion in $\sigma$, the previous expression for the wormhole with $\gamma=1$ reduces to
    \begin{equation}
    \tilde{\sigma}=\sigma+\frac{(5 \bar{r}-2)\left(2 \bar{r}^2-\bar{r}-2\right) \sigma^4}{4 \sqrt{\bar{r}-1} \bar{r}^7}+\mathcal{O}\left(\sigma^5\right)
    \label{eq:spincorrection1}
    \end{equation}

    In the case of the wormhole solution with $\gamma=2$, the power series expansion of the spin correction yields
    \begin{equation}
    \tilde{\sigma}=\sigma+\frac{\left(1-7 \bar{r}^2-3 \bar{r}^3+4 \bar{r}^4+2 \bar{r}^5\right)\sigma^4}{\bar{r}^{17 / 2} \sqrt{\bar{r}^2-1}}+\mathcal{O}\left(\sigma^5\right)
    \label{eq:spincorrection2}
    \end{equation}
    Hence, we will utilize the spin correction provided by $\tilde{\sigma} \neq \sigma$, as outlined in the previous two expressions, to recalculate the ISCO parameters once more; nevertheless, before doing so, let's see the form of the spin correction for the other relationship represented by TD-OKS SSC.
    \begin{figure*}
        \centering
    \includegraphics[width=1\linewidth]{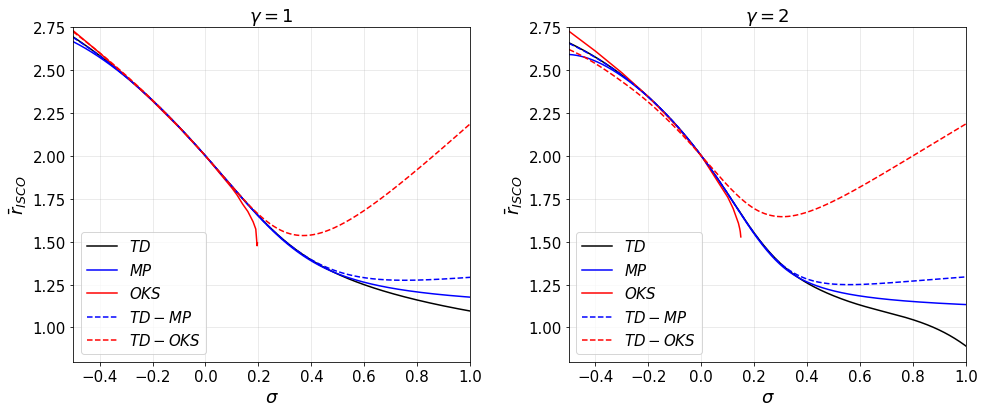}
    \caption{ISCO radius for each SSC compared to those corrected by taking $\tilde{\sigma} \neq \sigma$, for different values of spin $\sigma$ and for both cases $\gamma=1,2$.}
    \label{fig:fig6}
    \end{figure*}
    
    \begin{figure*}
    \centering
    \includegraphics[width=1\linewidth]{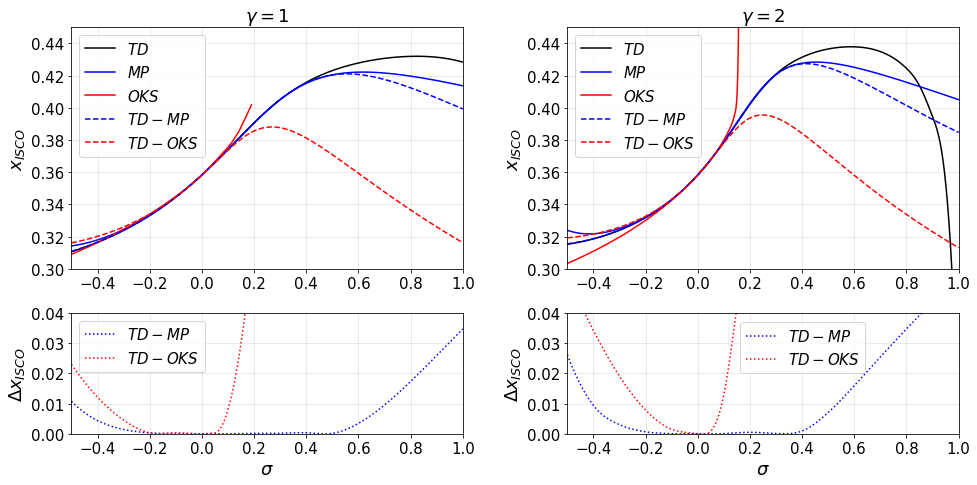}
    \caption{On top panels, we present the ISCO orbital frequency  parameter $x_{ISCO}$ for each SSC compared to those corrected by taking $\tilde{\sigma} \neq \sigma$. While in the lower panels, we show the relative differences $\Delta x_{ISCO}$ of the MP and OKS-SSCs with respect to the TD-MP and TD-OKS-SSCs corrections, using the Eqs.~\eqref{eq:relativediff2} and \eqref{eq:relativediff3}, respectively.  These quantities are given as a function of spin $\sigma$ and for both cases $\gamma=1,2$.}
    \label{fig:fig7}
    \end{figure*}

    \subsubsection{Spin Corrections for TD-OKS SCC}

    Because of the relation in Eq.~\eqref{eq:OKSmomentumvelocity}, we have $m=\mu$ in the OKS-SSC case, and using this result in Eq.~\eqref{eq:spincorrection}, it is possible to write the following equation in terms of the velocity,
    \begin{widetext}
    \begin{align}
    \label{spin_correction_TD_OKS}
        \tilde{\sigma}^2 = \sigma^2 + \delta r \Biggl \{ \Biggr.& g_{rr} \left[ (\sigma^{r\varphi})^2 \partial_r g_{\varphi \varphi}  + (\sigma^{tr})^2 \partial_r g_{tt} + \frac{2}{b_0}\left(u_t \sigma^{tr} - u_\varphi \sigma^{r\varphi}\right) \right] \notag \\
        &\Biggl. + \sigma^2 \left[ \frac{\partial_r g_{rr} }{g_{rr}} + \left( \frac{u_t}{g_{tt}} \right)^2 \partial_r g_{tt} + \left( \frac{u_\varphi}{g_{\varphi \varphi}} \right)^2 \partial_r g_{\varphi \varphi} \right] \Biggr\}
    \end{align}
    \end{widetext}
    where the normalized spin tensor is given again by Eq.~\eqref{eq:sigmaUpUp}. Using Eqs.~\eqref{eq:utorbitalfreq}, \eqref{eq:referenceVectorOKS1} and \eqref{eq:referenceVectorOKS2}, we obtain for the wormhole spacetime with $\gamma=1$ the following correction to the normalized spin
    \begin{equation}
    \tilde{\sigma}=\sigma + \frac{ \left(5 \bar{r}-2\right) \left(\bar{r}^2 +2\bar{r} - 2\right)}{2 \left(\bar{r}-1\right) \bar{r}^5} \sigma ^3 + \mathcal{O}\left(\sigma^4\right).
    \label{eq:spincorrection3}
    \end{equation}
    
    Similarly, the spin correction for the wormhole with $\gamma=2$ is given by the expression 
    \begin{widetext}
    \begin{equation}
        \tilde{\sigma}=\sigma
    + \frac{\left(5 \bar{r} - 2\right)}{2 \left(\bar{r}-1\right) \bar{r}^5} 
    \left[ \frac{\left(5 \bar{r} - 2\right)}{2}
    +\frac{\left(\bar{r} \left(\bar{r}^2+\bar{r}-2\right)-1\right)}{ \bar{r}^{1/2} \sqrt{\bar{r}+1}}\right]\sigma ^3 + \mathcal{O}\left(\sigma^4\right).
    \label{eq:spincorrection4}
    \end{equation}
    \end{widetext}
    
    We will now recalculate the ISCO parameters by applying the spin corrections provided by Eqs.~\eqref{eq:spincorrection1} and \eqref{eq:spincorrection2} for the TD-MP-SSC, and Eqs.~\eqref{eq:spincorrection3} and \eqref{eq:spincorrection4} for the TD-OKS-SSC, for the cases $\gamma=1$ and $\gamma=2$, respectively.

    Figure \ref{fig:fig6} displays the calculated ISCO radius $\bar{r}_{ISCO}$ for each SSC, as well as the $\bar{r}_{ISCO}$ derived by utilizing the spin correction. On the other hand, Figure \ref{fig:fig7} presents the ISCO orbital frequency parameter $x_{ISCO}$ for each SSC, as well as those corrected through the use of $\tilde{\sigma} \neq \sigma$, and the relative differences $\Delta x_{ISCO}$ between the MP- and OKS-SSCs with respect to the TD-MP and TD-OKS-SSCs corrections, as obtained through Eqs.~\eqref{eq:relativediff2} and \eqref{eq:relativediff3}, respectively. These plots demonstrate that the spin corrections to the ISCO parameters, $\bar{r}_{ISCO}$ and $x_{ISCO}$, based on $\tilde{\sigma}_{ISCO}$, effectively bring the TD-SSC results closer to the behavior of the MP-SSC. In particular, comparing $\Delta x_{ISCO}$ obtained in Fig.~\ref{fig:fig2} with those obtained by taking $\tilde{\sigma} \neq \sigma$ confirms that the results of TD-MP-SSC exhibit a behavior closer to that of the MP-SSCs, in contrast to the TD-SSC. However, the correction represented by TD-OKS-SSC does not exhibit such an improvement. In fact, it is possible to see more discrepancies than the case without the correction. 
    
    It is important to point out that the use of the spin correction once more allows for the calculation of the ISCO parameters for values of $\sigma$ where the computational routine had previously failed, without the spin correction. Therefore, it is clear that the spin corrections for the TD-MP and TD-OKS-SSCs effectively eliminate the divergences observed in both TD-SSC and OKS-SSC before the implementation of these corrections. Furthermore, by comparing the results obtained through the spin correction with those achieved previously using the centroid position correction, we can identify situations where one correction produces better results over the other for specific spin values. For instance, in the case of the TD-MP-SSC spin correction, smaller $\Delta x_{ISCO}$ values are obtained at $\sigma=0.3$, compared to those obtained through the centroid position correction presented in Fig.~\ref{fig:fig44}. However, for $\sigma=0.7$, the spin correction produces higher $\Delta x_{ISCO}$ values than those obtained through the centroid position correction; this confirms that the two corrections approach the ISCO parameters of each SSC differently. Moreover, as discussed in \cite{Timogiannis:2021ung}, this is because combining the two corrections (simultaneously) does not necessarily improve the approximation between the results of each SSC. Therefore, we will not combine the two corrections simultaneously in this study either.

%%%%%%%%%%%%%%%%%%%%%%%%%%%%%%%%%%%%%%%%%%%%%%%%%%%%%%%%%%%%%%%%%%%%%%%%%%%%%%%%%%%%%%%%%%%%%% Section 6 %%%%%%%%%%%%%%%%%%%%%%%%%%%%
%%%%%%%%%%%%%%%%%%%%%%%%%%%%%%%%%%%%%%%%%%%%%%%%%%%%%%%%%%%%%%%%%%
    \section{CONCLUSION} \label{sec:6}

    In this work, we compare different SSCs for spinning test particles moving in equatorial circular orbits around a Morris-Thorne traversable wormhole. We consider two wormhole solutions; $\gamma=1$ and $\gamma=2$, and the most known SSCs in the literature; i. e. TD-, MP-, and OKS-SSCs. 
    
    We begin by investigating the influence of each SSC on the particle's orbital frequency, $\hat{\Omega}$. To do so, we expand the orbital frequency in powers of the dimensionless particle's spin, $\sigma$; we carried out the expansion up to the third-order, where differences in all the SSCs started to appear. As expected, our results show that the zero-order frequency is the same for all SSCs; this frequency corresponds to well-known Keplerian frequency $1/\overline{r}^\frac{3}{2}$, see Table.~\ref{table:Frequency01} and \ref{table:Frequency02}, for $\gamma=1$ and $\gamma=2$, respectively. Moreover, the equivalence in the orbital frequency extends to the first order in both wormhole solutions; nevertheless, we start seeing some differences when considering the second order of approximation. For example, we found that OKS-SSC begins to differ from the TD- and MP-SSCs, which still have the same behavior at this order of magnitude. Then, when we considered the third order of approximation, we found that all the SSCs differ. The same conclusion appears for the Schwarzschild and Kerr spacetimes~\cite{Timogiannis:2021ung, Timogiannis:2022bks}. Therefore, the fact that the TD-SSC is more compatible with the MP-SSC than the OKS-SSC also extends to the Morris-Thorne traversable wormhole.
    
    In the case of the ISCO, we found that it decreases as the particle's spin increases; this is the general feature in all SSCs and for both wormhole solutions considered in this paper. Moreover, although $r_{ISCO}$ has the same behavior in the vicinity of $\sigma=0$, it is worth noticing that, in the OKS-SSC, the ISCO diverges at some particular value of $\sigma$; for example, when $\gamma=1$, the ISCO radius diverges at $\sigma\approx 0.2$. A similar behavior occurs at $\sigma\approx0.15$ when $\gamma=1$ (note that increasing the gamma shifts the limit value to the left). This behavior is mainly the consequence of choosing a specific SSC. Recall that, from the physical point of view, a different SSC corresponds to a different location of the particle's centroid. 

    On the other hand, according to the relative difference with respect to the TD-SSC, our results show that the MP-SSC behaves similarly to TD-SSC when the particle's spin belongs to the interval $-0.2<\sigma<0.4$ of the wormhole solution with $\gamma=1$. This interval reduces when $\gamma=2$; in that case, the MP-SSC behaves similarly to the TD-SSC if $-0.2<\gamma<0.3$, where $\Delta x_{ISCO}<<0.01$. In the case of the OKS-SSC with $\gamma=1$, its behavior is similar to TD-SSC in the interval $-0.4<\sigma<0.0$ (with relative differences smaller than $0.005$). When $\gamma=2$, the OKS-SSC is similar to TD-SSC only in the region near $\sigma=0$. Therefore, the wormhole parameter $\gamma$ does influence the ISCO radius depending on the SSC. In particular, when spinning test particles move with $\sigma>0$.  

    Through the particles' centroid corrections, it is possible to explain (up to some degree) the differences between each SSC. Therefore, to see how close each SSC is to the other, we investigate (separately) the radial and spin corrections taking as a reference the TD-SSC. Our results show that, in the case of the wormhole solution with $\gamma=1$, the TD-SSC behaves very closely to the MP-SSC after the radial correction; in this case, the values of $\Delta x_{ISCO}$ are smaller for a longer interval, even for positive values of the particle's spin $\sigma$. Nevertheless, the situation is different for the wormhole solution with $\gamma=2$; although there is an improvement in the convergence, the correction of the TD-SSC to the MP-SSC is not as good as the wormhole solution with $\gamma=1$. On the other hand, the radial correction of the TD-SSC to OKS-SSC shows an improvement in the interval $-0.2<\sigma<0.2$ for both wormhole solutions in contrast to the TD-SSC without correction. Hence, the wormhole parameter $\gamma$ influence on the radial corrections.   

    As mentioned above, the ISCO radius diverges for some value of $\sigma$ (depending on the value of the wormhole parameter $\gamma$) in the case of the OKS-SSC. However, when considering the spin correction in the TD-SSC, the divergence disappears independently of the wormhole solution. Similarly, the divergence of $\Delta x_{ISCO}$ found for the solution with $\gamma=2$ in the TD-SSC vanishes when one considers the spin corrections of the MP-SSC.
    
    Finally, from the algorithm proposed in \cite{Timogiannis:2021ung} to investigate the orbital frequencies in the three SSCs, it is clear that the differences are connected deeply with the radial correction given in Eq.~\eqref{s5e3} since the spin tensor, $S^{\mu\nu}$, depends on the reference four-vector $V^\mu$. In this sense, an equatorial circular orbit may degenerate into a non-circular one when changing from one SSC to another; this is why the SSCs become different at some order of magnitude (up to the third order for Schwarzschild and Kerr black holes and Morris-Thorne traversable wormholes). However, the fact that the ISCO is a special limit and that additional improvements such as the spin corrections in Eqs.~\eqref{spin_correction_TD_MP} and \eqref{spin_correction_TD_OKS} were not enough to explain the differences at the ISCO, also suggest that the pole-dipole approximation does not work anymore in curved spacetime and becomes necessary to include higher order terms in the multipole expansion.
%%%%%%%%%%%%%%%%%%%%%%%%%%%%%%%%%%%%%%%%%%%%%%%%%%%%%%%%%%%%%%
    \section*{Acknowledgements} \label{sec:acknowledgements}
    This work was supported by the Universidad Nacional de Colombia, Hermes Grant Code 57057, and by the Research Incubator No.64 on Computational Astrophysics of the Observatorio Astronómico Nacional. C.A.B.G. acknowledge the support of the Ministry of Science and Technology of China (grant No.~2020SKA0110201) and the National Science Foundation of China (grants No.~11835009). 

%%%%%%%%%%%%%%%%%%%%%%%%%%%%%%%%%%%%%%%%%%%%%%%%%%%%%%%%%%%%%%%%%%%%%%%%%%%%%%%%%%%%%%%%%%%% Bibliography %%%%%%%%%%%%%%%%%%%%%%%%%%%%
%%%%%%%%%%%%%%%%%%%%%%%%%%%%%%%%%%%%%%%%%%%%%%%%%%%%%%%%%%%%%%%%%%%

 %%%%%%%%%%%%%%%%%%%%Appendix%%%%%%%%%%%%%%%%%%%%%%%%%%%%%%%%%%%%%%

    \begin{widetext}
    \section{Equations for $\gamma =2 $ using the TD-SSC} \label{sec:appendixTD}
    In the case of $\gamma=2$, $\rho_2$, $\rho_1$, and $\rho_0$ takes the form
    \begin{equation}
        \label{appAe1}
        \begin{aligned}
        \rho_2&=r e^{-\frac{2 b_0}{r}} \left(b_0^2 S^2-\mu ^2 r^4\right),\\
        %%%%%%%%
        \rho_1&=\frac{b_0 \mu  S e^{-\frac{3 b_0}{r}} \left(3 b_0^2 r-b_0^3-2 r^3\right)}{\sqrt{r^2-b_0^2}},\\
        %%%%%%%%
        \rho_0&=\frac{b_0 e^{-\frac{4 b_0}{r}} \left(b_0 S^2 \left(-b_0 r^2-3 b_0^2 r+b_0^3+2 r^3\right)+\mu ^2 r^6\right)}{r^4},\\
        \end{aligned}
        \end{equation}
     The four-momentum components, $p^t$ and $p^{\varphi}$, of the spinning test particle under the TD-SSC for $\gamma=2$ are
         \begin{equation}
            \begin{aligned}
            p^\varphi&=\frac{\mu }{\sqrt{e^{\frac{2 b_0}{r}}\mathcal{F}^2_2(\Omega,r;S)-r^2}},\\\\
            %%%%%%%%%%
            p^t&=-\frac{\mu  e^{\frac{2 b_0}{r}} \mathcal{F}_2(\Omega,r;S)}{ \sqrt{e^{\frac{2 b_0}{r}}\mathcal{F}^2_2(\Omega,r;S)-r^2}}.
            \end{aligned}
         \end{equation} 
    Where we define
    \begin{equation}
    \mathcal{F}_2(\Omega,r;S)=\frac{\mathcal{A}_2(\Omega,r;S)}{b_0 \left(b_0^2 \mu +b_0 S \Omega  e^{\frac{b_0}{r}} \sqrt{r^2-b_0^2}-\mu  r^2\right)},
     \end{equation}
     and,
    \begin{equation}
         \mathcal{A}_2(\Omega,r;S)=\left(\mu  r^3 \Omega  \left(r^2-b_0^2\right)+\frac{b_0 S e^{-\frac{b_0}{r}} \sqrt{r^2-b_0^2} \left(-b_0 r^2-3 b_0^2 r+b_0^3+2 r^3\right)}{r^2}\right).
     \end{equation}
    %\end{widetext}
    %\begin{widetext}
    
    \section{Equations for $\gamma =2 $ using the MP-SSC} \label{sec:appendixMP}
    The four-momentum components of the spinning test particle under the MP-SSC and for the wormhole solution with $\gamma=2$ are 
    \begin{equation}
    p^{t}=\frac{e^{\frac{b_0}{r}}\left[-e^{\frac{b_0}{r}} S \Omega b_0 \sqrt{r^2-b_0^2}+r^2\left(m-e^{\frac{2 b_0}{r}} m r^2 \Omega^2+e^{\frac{3 b_0}{r}} r S \Omega^3 \sqrt{r^2-b_0^2}\right)\right]}{r^2\left(1-e^{\frac{2 b_0}{r}} r^2 \Omega^2\right)^{3 / 2}}
    \end{equation}
    \begin{equation}
    p^{\varphi}=\frac{e^{\frac{b_0}{r}} r^3 \Omega\left[m r\left(1-e^{\frac{2 b_0}{r}} r^2 \Omega^2\right)+e^{\frac{b_0}{r}} S \Omega \sqrt{r^2-b_0^2}\right]-S b_0 \sqrt{r^2-b_0^2}}{r^4\left(1-e^{\frac{2 b_0}{r}} r^2 \Omega^2\right)^{3 / 2}}
    \end{equation}
    %\end{widetext}
    For the quartic Eq.~\eqref{eq:MPOmegaPoly}, the polynomial coefficients for the orbital frequency for $\gamma=2$ are
    \begin{equation}
    \begin{aligned}
    \xi_4=& e^{\frac{4 b_0}{r}} m r^5\left(r+b_0\right), \\
     \xi_3=&-e^{\frac{3 b_0}{r}} S \sqrt{r^2-b_0^2}\left(r^3-2 r^2 b_0-3 r b_0^2+ b_0^3\right),  \\
      \xi_2=&- e^{\frac{2 b_0}{r}} m r^2\left(r+b_0\right)^2, \\
       \xi_1=&- e^{\frac{b_0}{r}} S b_0 \left(r+2 b_0\right) \sqrt{1-\frac{b_0^2}{r^2}},  \\
        \xi_0=& m b_0 \left(r+b_0\right).  \\
    \end{aligned}
    \end{equation}
    %\end{widetext}
    
    \section{Equation for the Orbital Frequency using the OKS SSC} \label{sec:appendix}
   
   In order to obtain the equation that will give the orbital frequency using the OKS SSC, we replace the momentum given by Eq.~ (\ref{eq:OKSmomentumvelocity}) and the components of the spin tensor of Eq.~(\ref{eq:SpinTensorComponents}), in the MPD Eqs.  (\ref{eq:MPD1}) and (\ref{eq:MPD2}), to obtain
   %\begin{widetext}
   \begin{align}
       m u^t \left( \Gamma^r_{tt} + \Gamma^r_{\varphi \varphi} \Omega^2 \right) = & \left( R^r_{ttr} g_{\varphi \varphi} w^\varphi  + R^r_{\varphi r \varphi} \Omega g_{tt} w^t \right) \sqrt{\frac{g_{\theta \theta}}{g}} s \\
       \Gamma^t_{tr} g_{tt} w^t = &- \Gamma^\varphi_{r\varphi} g_{\varphi \varphi} w^\varphi .
   \end{align}
    %\end{widetext}

    Using the components of the reference vector of Eqs.~ (\ref{eq:referenceVectorOKS1}) and (\ref{eq:referenceVectorOKS2}) together with the relation of Eq.~ (\ref{eq:utorbitalfreq}), we transform these equations into the following 6th-degree polynomial for the orbital frequency of the spinning test particle,
    \begin{equation}
        \psi_6 \Omega^6 + \psi_4 \Omega^4 + \psi_2 \Omega^2 + \psi_0 =0 \label{eq:OKSOmegaPoly}
    \end{equation}
    where 
    %\begin{widetext}
            \begin{align}
            \psi_0 = & \left( R^r_{ttr} \right)^2 \frac{g_{tt}}{g_{tt}} s^2 + g_{tt} \left( \Gamma^r_{tt} \right)^2 m^2\\
            \psi_2 = & \left( R^r_{ttr}  g_{\varphi \varphi} - 2 R^r_{\varphi r \varphi} \frac{\Gamma^\varphi_{r \varphi}}{\Gamma^t{tr}}  g_{tt}  \right) \frac{R^r_{ttr}}{g_{rr}} s^2 + \left(  \left( \frac{\Gamma^\varphi_{r\varphi}}{\Gamma^t_{tr}} \right)^2 \Gamma^r_{tt} g_{\varphi\varphi} + 2 \Gamma^r_{\varphi\varphi} g_{tt} \right)\Gamma^r_{tt} m^2 \\
            \psi_4 = & \left( R^r_{\varphi r \varphi} \frac{\Gamma^\varphi_{r \varphi}}{\Gamma^t{tr}} g_{tt} - 2 R^r_{ttr} g_{\varphi \varphi}  \right) \frac{\Gamma^\varphi_{r \varphi}}{\Gamma^t{tr}} \frac{R^r_{\varphi r \varphi}}{g_{rr}} s^2 + \left( 2 \left( \frac{\Gamma^\varphi_{r\varphi}}{\Gamma^t_{tr}} \right)^2 \Gamma^r_{tt} g_{\varphi\varphi} + \Gamma^r_{\varphi\varphi} g_{tt} \right)\Gamma^r_{\varphi\varphi} m^2\\
            \psi_6 =& \left[\left( R^r_{\varphi r \varphi} \frac{\Gamma^\varphi_{r \varphi}}{\Gamma^t{tr}} \right)^2 \frac{s^2}{g_{rr}} + \left( \frac{\Gamma^\varphi_{r\varphi} \Gamma^r_{}\varphi \varphi}{\Gamma^t_{tr}} \right)^2 m^2 \right] g_{\varphi \varphi}.
    \end{align}        
    \end{widetext}
	\end{document}